\RequirePackage{fix-cm}
\documentclass[twocolumn,epjc3]{svjour3}  
\smartqed  % flush right qed marks, e.g. at end of proof
\RequirePackage{graphicx}

\usepackage[british]{babel}
\usepackage{hyperref}
\usepackage{graphicx}
\usepackage{cuted}
\usepackage{caption}
\usepackage{subcaption}
\usepackage{dblfloatfix}  
\usepackage[separate-uncertainty,bracket-numbers = false,
  table-align-uncertainty = true, table-align-exponent=false, 
exponent-product=\cdot,parse-numbers = false]{siunitx}

\catcode`\@=11 % @ signs are now treated as letters
\def\parenbar{\mathpalette\p@renb@r}
\def\p@renb@r#1#2{\vbox{%
		\ifx#1\scriptscriptstyle \dimen@.7em\dimen@ii.2em\else
		\ifx#1\scriptstyle \dimen@.8em\dimen@ii.25em\else
		\dimen@1em\dimen@ii.4em\fi\fi \offinterlineskip
		\ialign{\hfill##\hfill\cr
			\vbox{\hrule width\dimen@ii}\cr
			\noalign{\vskip-.3ex}%
			\hbox to\dimen@{$\mathchar300\hfil\mathchar301$}\cr
			\noalign{\vskip-.3ex}%
			$#1#2$\cr}}}
\catcode`\@=12 % @ signs are no longer letters
\def\nuan{\ensuremath{\parenbar{\nu}\kern-0.4ex}}

\addto\extrasbritish{

}

\def\Pi{\ensuremath{\pi^{\pm}}}
\def\PiP{\ensuremath{\pi^{+}}}
\def\PiM{\ensuremath{\pi^{-}}}
\def\Mu{\ensuremath{\mu^{\pm}}}
\def\Nu{\ensuremath{\nu}}
\def\MuP{\ensuremath{\mu^{+}}}
\def\Kp{\ensuremath{K^{+}}}
\def\NuANuMu{\ensuremath{\nuan_\mu}}
\def\NuANuE{\ensuremath{\nuan_e}}
\def\NuANuTau{\ensuremath{\nuan_\tau}}
\def\NuE{\ensuremath{\nu_e}}
\def\NuMu{\ensuremath{\nu_\mu}}

\def\ANuMu{\ensuremath{\bar\nu_\mu}}
\def\ANuE{\ensuremath{\bar\nu_e}}
\def\PiMuNu{\ensuremath{\Pi \to \Mu \NuANuMu}}
\def\KMuNu{\ensuremath{\Kp \to \MuP \NuMu}}
\def\tagging{{tagging}}
\def\tagged{{tagged}}
\def\interacting{{interacting}}
\newcommand{\Oo}[1]{\ensuremath{\mathcal{O}(#1)}}
\newcommand{\dmsq}[1]{\rm \Delta m^2_{#1}}

\journalname{Eur. Phys. J. C}

\begin{document}
\sloppy %avoid text going over the column end
\title{Neutrino Tagging: a new tool for accelerator based neutrino experiments}

\titlerunning{Neutrino Tagging}        % if too long for running head

\author{Mathieu Perrin-Terrin\thanksref{g1,cppm}
}

\thankstext{g1}{ANR-19-CE31-0009}
% \thankstext{e1}{e-mail: fauthor@example.com}


\institute{Aix Marseille Univ, CNRS/IN2P3, CPPM, Marseille, France \label{cppm}
}

\date{Received: date / Accepted: date}

\maketitle

\begin{abstract}

This article describes a new experimental method for accelerator based neutrino experiments called neutrino \textit{tagging}. The method consists in exploiting the neutrino production mechanism, the \PiMuNu\ decay, to kinematically reconstruct the neutrino properties from the decay incoming and outgoing charged particles. The reconstruction of these particles relies on the recent progress and on-going developments in silicon particle detector technology. A detailed description of the method and achievable key performances is presented, together with its potential benefits for short and long baseline experiments. Then, a novel configuration for long baseline experiments is discussed in which a tagged beam would be employed together with mega-ton scale natural deep water Cherenkov detectors. The coarseness of this type of detectors is overcome by the precision of the tagging and, conversely, the rate limitation imposed by the tagging is outweighed by the size of the detector. These mutual benefits result in an affordable design for next generations of long baseline experiments. The physics potential of such experiments is quantified using the Protvino to KM3NeT/ORCA setup as a case study for which an unprecedented sensitivity to the leptonic CP violation could be achieved.

\end{abstract}

% \newpage
% \onecolumn
% \hypersetup{hidelinks}
% \setcounter{tocdepth}{2}
% \tableofcontents
% \newpage
\section{Introduction}

The discovery of neutrino oscillation with atmospheric neutrinos and its confirmation with solar neutrinos have inaugurated a rich field in fundamental physics. Following these first measurements, new 
experiments were designed to operate with controlled neutrino sources to more precisely measure the neutrino oscillation parameters. In this context, experiments have been performed with neutrino beams with energies of \SI{\mathcal{O}(1-10)}{GeV} produced at particle accelerators. This type of experiments typically requires two neutrino detectors. The first one is installed near the accelerator to characterise the initial neutrino flux. The second one is placed further downstream and measures the flux after oscillation. For studying the standard neutrino oscillation, the distance over which the neutrinos propagate has to be \SI{\mathcal{O}(100-1000)}{km}. Hence, these setups are referred to as long baseline neutrino experiments (LBNE's) in contrast to short baseline neutrino experiments (SBNE's) for which the distance of propagation is \SI{\mathcal{O}(10-100)}{m}. The physics case of the latter is the study of non-standard neutrino oscillations, neutrino cross-sections and interactions.

% 
% In the two-flavour approximation, the oscillation is proportional to $\sin^2\left( 1.27 \cdot\Delta m^2\cdot L / E \right)$ where $\Delta m^2$ is the squared mass difference between the two neutrino mass eigenstates in \si{eV^2},  $L$ is the propagation distance, or \textit{baseline} in \si{km} and $E$ the neutrino energy in \si{GeV}. The typical energies for neutrinos produced at accelerators are \SI{\mathcal{O}(1)}{GeV}. Hence for largest squared mass differences, $\Delta^2m_{31}$ and $\Delta^2m_{32}$, the first oscillation maximum is reached at a baseline of about \SI{500}{km}.

% which allows to study the 
% The oscillation 
% 
% 
% The distance between the place where the neutrinos are produced and the far detector, called \textit{baseline}, depends on the neutrino energy. For neutrino energies of about \SI{\mathcal{O}(1)}{GeV} the standard neutrino oscillation, the baseline is typically few hundred kilometers.

The first generations of LBNE's, K2K, MINOS, T2K, NOVA, have successfully improved the knowledge on the mixing angles and the squared mass splittings. The next generation of experiments, DUNE~\cite{hep-ph_DUNE_2020,hep-ph_DUNE_2020a,hep-ph_DUNE_2020b} and T2HK~\cite{hep-ph_Hyper-Kamiokande_2018,hep-ph_Hyper-Kamiokande_2021}, are being constructed to determine the neutrino mass ordering and to study CP violation in the neutrino sector. These new experiments rely on the same methodology but employ larger detectors and more powerful beams to collect larger neutrino samples. Moreover, they implement new techniques such as movable near detectors to better characterise the neutrino flux and so reduce the systematic uncertainties. In parallel to these LBNE's, several SBNE's, have been carried out: LSND, MiniBooNE and MicroBooNE. These experiments have indicated anomalous oscillation patterns which, as of today, still remain puzzling.

In all these SBNE's and LBNE's, the properties of the neutrinos are obtained based solely on the neutrino interaction final state. This article proposes a new method to refine the measurement of these properties by also exploiting the neutrino production mechanism, the \PiMuNu\ decay. The principles of the method are described in~\autoref{sec:1}. In~\autoref{sec:2}, a generic experimental setup using this method is presented together with estimates for the most relevant technical performances. In \autoref{sec:3}, this generic design is applied to the case of a LBNE and preliminary sensitivity estimates to key observables are presented to illustrate the physics potential of the setup. Finally, summary and prospects are discussed in \autoref{sec:conclusion}.

\section{The Neutrino Tagging Method}
\label{sec:1}

\subsection{Conceptual Description}

The neutrino beams produced at accelerators are primarily obtained by generating an intense beam of pions that decay in flight as \PiMuNu. The possibility of extracting useful information from the decay has been identified early~\cite{hep-ph_Pontecorvo_1979,hep-ph_Nedyalkov_1984,hep-ph_Bohm_1987,hep-ph_BernsteinEtAl_1990} but never completely implemented as proposed in this article\footnote{At Protvino\cite{hep-ph_AnikeevEtAl_1998}, few interactions of neutrinos from \PiMuNu\ were associated with the \Mu\ from the decay.}.
Continuous progress in silicon pixel detectors~\cite{hep-ph_AglieriRinellaEtAl_2019,hep-ph_Lai_2018,hep-ph_SadrozinskiEtAl_2013} allows to operate beam trackers at increasingly high particle rates such that a neutrino beam line instrumented with silicon trackers becomes conceivable.
These instruments would allow to reconstruct all \PiMuNu\ decays from the tracks of the incoming \Pi\ and outgoing \Mu. Using this information, for each decay, a \textit{tagged} neutrino could be formed with the following properties:
\begin{itemize}
	\item a muonic initial neutrino flavour, to match the charged lepton one,
	\item a chirality opposite to the lepton one, or deduced from the pion electric charge,
	\item a direction and energy fulfilling momentum and energy conservation at the decay.
\end{itemize}

Based on time and angular coincidence, each neutrino \interacting\ in the detector could be associated with a single tagged neutrino.
% Then, a one-to-one association between the \textit{interacting} neutrinos and \textit{tagged} neutrinos could be made using time and space coincidence.
The resulting \textit{associated} neutrino sample would allow to access to a rich physics program, as described in the next section.

\subsection{Expected Benefits}
\label{sec:benefits}

The neutrino \textit{tagging} technique has three main advantages. First, it enables the reconstruction of nearly all neutrinos in the beam. Second, it allows to track each \interacting\ neutrino from the detection back to the production at the \PiMuNu\ decay. This ability allows in turn to precisely reconstruct the \interacting\ neutrino properties by exploiting the decay kinematics. These advantages enter in numerous ways into the study of neutrino physics as described in the next paragraphs.

% by improving the neutrino flux knowledge, reducing background for new physics search, improving the energy reconstruction and the neutrino flavour identification, refining the cross-section measurement, determining the neutrin chirality.

% By allowing to reconstruct nearly all neutrinos in a beam, by enabling the individual tracking of each \interacting\ neutrino from creation to detection, and, by providing an unprecedented level of precision for the neutrino energy reconstruction, the \tagging\ technique opens possibilities that could not be envisaged so far.

\paragraph{Improved flux measurement}

% The ability to reconstruct all neutrinos in a beam allows to completely characterise the flux in terms of energy, chirality and initial flavour at any place downstream of the beam. This nearly perfect knowledge of the flux would greatly improve neutrino oscillation measurements.
% 
% Indeed, these measurements normally proceed by comparing the flux composition at a short distance from the neutrino generation region, where the neutrino oscillation has not yet occured, and, further downstream, at a distance corresponding to one of the oscillation maxima.
% As the two locations are separated by distances up to of several hundreds or thousands of kilometers, the two detectors employed to measure the flux are covering very different solid angles. This difference requires to apply corrections to the measured fluxes.
% These corrections are not trivial as the mean neutrino energy depends on the neutrino direction, and, the neutrino cross-section depends on the energy.
% As a result, they induce large systematic uncertainties on the neutrino oscillation parameters. With a \tagged\ experiment, these corrections would not be needed and the related systematic uncertainties would be removed.

The ability to precisely and individually reconstruct all beam neutrinos from \PiMuNu\ decays is very useful to determine the neutrino flux and its composition in terms of energy, flavour and chirality.

At SBNE's, the angular resolution on the \tagged\ neutrinos is sufficient to perfectly determine the neutrino flux, \textit{i.e.} predict individually which beam neutrino from \PiMuNu's are in the detector acceptance. Such a perfect flux determination is extremely useful to measure neutrino cross sections.

At LBLNE's, the angular resolution might not be sufficient to predict, event-by-event, which \tagged\ neutrinos are in the detector acceptance. However, the \tagging\ provides stringent constraints on the ratio of the flux at the far and near detectors. This ratio is a significant source of uncertainties for oscillation studies~\cite{hep-ph_HuberEtAl_2008,hep-ph_BrancaEtAl_2021}.

\paragraph{Background suppression} % add also Huber paper here

One of the main backgrounds for the studies of neutrino oscillation in the appearance $\NuANuMu \to \NuANuE$ channels at SBNE and LBNE are the non-oscillated \NuANuE\ beam components~\cite{hep-ph_LSND_2001,hep-ph_MiniBooNE_2013,hep-ph_HuberEtAl_2008}. The \tagging\ technique would allow to significantly reduce this background~\cite{hep-ph_LudoviciEtAl_1996} as the non-oscillated \interacting\ \NuANuE\ will not coincide with any \tagged\ \NuANuMu\ and could thus be discarded.

% \paragraph{Background suppression for New Physics searches}
% New Physics can be searched for at SBNE by looking for anomalous $\NuANuMu \to \NuANuE$ events~\cite{hep-ph_LSND_2001,hep-ph_MiniBooNE_2013}. One of the background for these searches are the non oscillated \NuANuE\ beam component. The tagging technique would allow to significantly reduce this background~\cite{hep-ph_LudoviciEtAl_1996} as the non oscillated \interacting\ \NuANuE\ will not coincide with any \tagged\ \NuANuMu\ and could thus be discarded.

% At short base line neutrino experiment (SBNE), an individual tracking of the neutrino from production to detection, would allow to claim a significant evidence for new physics, or set strong constraints on the parameter phase space, from a much smaller number of anomalous events in the $\NuANuMu \to \NuANuE$ than in a conventional experiment. Indeed, the \tagging\ technique would allow to discard a large fraction of background events, i.e. events orignating from unoscillated \NuANuE, as for these events the \interacting\ \NuANuE\ will not coincide with any \tagged\ \NuANuMu.
% it is 90%
% is a game changer for neutrino physics. oscillation systematics but also cross section.

% Flux... how many neutrino of which energy chirality and inital flavour crossed the detector.
\paragraph{Improved energy reconstruction}

The \textit{tagged} neutrino energy measurement is expected to largely surpass the methods relying on the neutrino interaction. To illustrate this, one can consider the most forward neutrinos which are very relevant for on-axis LBNE's. These neutrinos have an energy, $E_\nu$, equal to
\begin{eqnarray}
	E_\nu  &=& (1-m^2_\mu/m^2_\pi) \cdot E_\pi \nonumber\\
	&=&0.43 \cdot E_\pi,
	\label{eq:}
\end{eqnarray}
where $m_\mu$ and $m_\pi$ are the \Mu\ and \Pi\ masses and $E_\pi$ the \Pi\ energy. Hence, the $E_\nu$ resolution is equal to the \Pi\ energy resolution which, in the ultrarelativitic hypothesis, is equal to the \Pi\ momentum resolution. A magnetic spectrometer can easily provide \Oo{0.1-1}\% precision for \Pi's with momenta of \SI{\Oo{1-10}}{GeV/c}~\cite{hep-ph_AglieriRinellaEtAl_2019,hep-ph_BoothEtAl_2019} with almost no uncertainties on the energy scale. By contrast, the reconstruction of the energy from the neutrino interaction final state is much more challenging. For instance, in a charged current (CC) interaction, the charged lepton recoils against an hadronic system which is subject to substantial stochastic fluctuations. These fluctuations induce variations of the light yield at Cherenkov water detectors~\cite{hep-ph_Adrian-MartinezEtAl_2017}, and of the ionisation charge at liquid argon detectors~\cite{hep-ph_FriedlandEtAl_2019}.
% As an example, the reconstruction of neutral current (NC) interactions or \NuANuTau\ charge current (CC) interactions are strongly affected by the outgoing neutrinos which reduce the visible energy by an unknown amount. 
% 
As a result, the resolutions obtained with these detectors are about one order of magnitude worse than the one expected with the \tagging\ technique. Moreover, relating the light yield or the ionisation charge to the neutrino energy relies on interaction models which induce significant uncertainties on the energy scales. These effects are notably detrimental to the study of neutrino oscillations~\cite{hep-ph_AnkowskiEtAl_2015,hep-ph_CLASe4v_2021}.

Hence, the \tagging\ technique allows to further reduce systematic uncertainties related to energy reconstruction and opens new possibilities to resolve the energy dependent patterns of the neutrino oscillation. This ability will further help to reduce the impact of systematic uncertainties as pointed in the conclusions of~\cite{hep-ph_HuberEtAl_2008}.

\paragraph{Improved neutrino flavour identification}

The \tagged\ neutrino energy reconstruction is independent of the neutrino interaction final state. Hence, by comparing the \tagged\ neutrino energy to the visible energy deposited in the detector by the \interacting\ neutrino, one could determine the process undergone by the neutrino during the interaction.

For example, this ability would allow to identify NC events as they release a smaller visible energy than CC interactions due to the outgoing neutrinos. The rate and spectrum of the NC events are unaffected by the neutrino oscillation. Hence they are conventionally considered as a background for the oscillating signal. However, in a \tagged\ experiment, not only these events could be isolated from the signal, but they could also serve the analysis for instance to further constrain the neutrino flux.

Similarly, \NuANuTau's undergoing CC interaction release a smaller visible energy than \NuANuMu-CC's or \NuANuE-CC's due to the neutrinos produced by the $\tau^\pm$ decay. Hence, the same technique could be used to select a neutrino sample enriched in \NuANuTau. Such a sample would be extremely valuable as the \NuANuTau\ appearance channel is essential to constrain the oscillation matrix unitarity~\cite{hep-ph_ParkeEtAl_2016}.

\paragraph{Improved neutrino interaction modeling}
The precise flux determination and energy reconstruction provided by the \tagging\ would allow to improve cross-section measurements, which will be very important for the next generations of LBNE's~\cite{hep-ph_HuberEtAl_2008}.
Indeed, the nearly perfect knowledge of the flux would allow to reduce the uncertainties on \NuANuMu\ absolute cross sections but also the energy dependence.

Moreover, as \tagged\ neutrinos are reconstructed independently of the neutrino interaction final state, they are excellent probes to refine the phenomenological models used to infer the neutrino energy from the neutrino–nucleus interactions~\cite{hep-ph_AnkowskiEtAl_2015,hep-ph_CLASe4v_2021}.

\paragraph{Event by event chirality determination}

The \tagging\ technique allows to determine event-by-event the neutrino chirality. Hence in a \tagged\ neutrino experiment, the alternation of the beam polarity is no longer needed and both neutrinos and anti-neutrinos can be collected together. This ability allows to collect data samples twice as large as the ones that a conventional beam experiment would for the same beam power and data taking period. Moreover, collecting both chiralities together is a strong asset to further reduce systematic uncertainties in the attempt to precisely determine the leptonic CP violating phase for which, the asymmetry between neutrinos and anti-neutrinos is crucial.

\section{Experimental Setup Design}
\label{sec:2}
The two keystones of the \tagging\ technique are the abilities to track all charged particles in a neutrino beam line and to associate the \interacting\ neutrinos to the \tagged\ ones. The following paragraphs describe how these two challenges can be addressed.

\begin{figure*}[b]% to keep image and caption on one page
		\includegraphics[width=\textwidth]{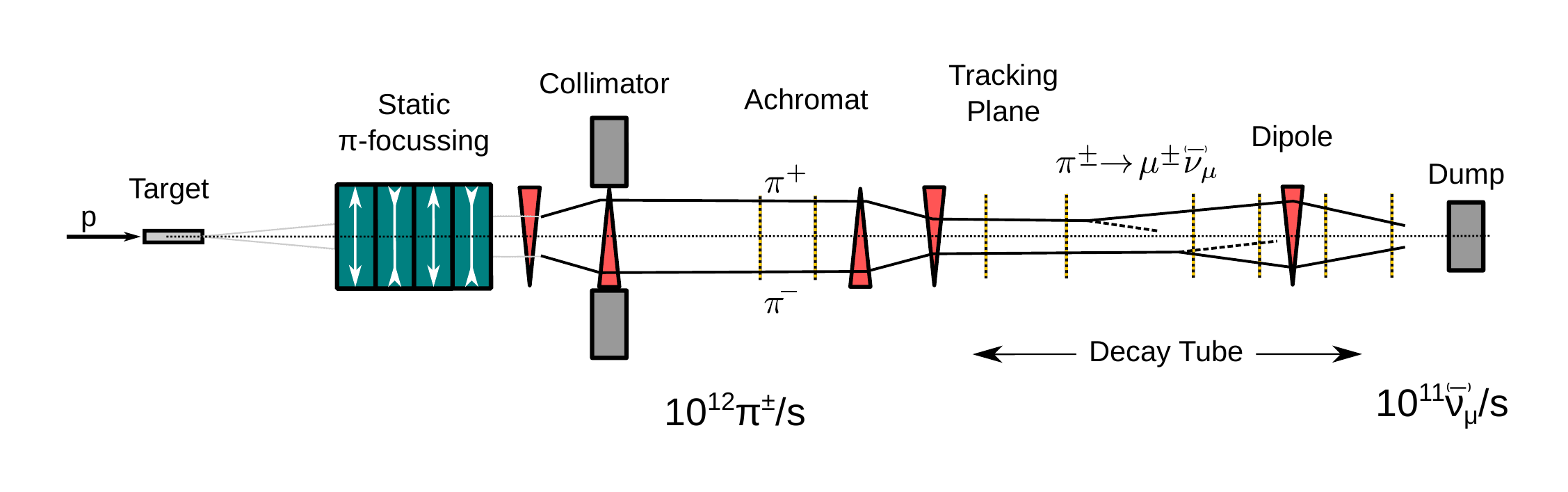}
	\caption{Schematic of a possible beam line enabling neutrino \tagging. Blue rectangles represent quadrupoles, red triangles dipoles and vertical dotted lines correspond to tracking planes. The number and location of those are not optimised. The schematic is described in more details in the text.}\label{fig:beamline}%      only if needed  
\end{figure*}

\subsection{Tracking Capabilities}
\label{sec:1.1}

Until recently, the use of silicon pixel trackers inside a neutrino beam line was prevented by the very high particle rates of these environments. In the past few years, significant progress has been achieved to increase the particle rate at which these instruments can be operated. In particular, the NA62 Collaboration has developed a beam tracker, called GigaTracKer~\cite{hep-ph_NA62Collaboration_2010,hep-ph_NA62_2017c,hep-ph_AglieriRinellaEtAl_2019}, able to withstand an instantaneous hadron rate of \SI{750\cdot{10^6}}{particle/s} with a peak flux of \SI{2.0\cdot{10^6}}{particle/s/mm^2}. The particle tracking at these rates is possible thanks to time-resolved pixels with a resolution of \SI{130}{ps}. The detector life time in this environment is limited and expected to correspond to a total integrated fluence normalised, under the Non Ionizing Energy Loss (NIEL) scaling hypothesis, to a \SI{1}{MeV} neutron equivalent fluence of \SI{\mathcal{O}({10^{14-15})}}{1~MeV~n_{eq}/cm^2}.
The LHC experiments have started to develop similar detectors~\cite{hep-ph_LHCbCollaboration_2017,hep-ph_LHCbVELOGroup_2021} for the high luminosity LHC (HL-LHC) upgrade foreseen for 2028. These detectors should be able to operate at even higher particle rates. They foresee a maximum flux of \SI{20\cdot10^6}{particle/s/mm^2}, a total fluence of \SI{\mathcal{O}({10^{16-17})}}{1~MeV~n_{eq}/cm^2} and a hit time resolution of \SI{30}{ps}.
Assuming that the beam particles are spread over \SI{\mathcal{O}({0.1})}{m^2}, these devices would allow to track a beam with a rate of \SI{\mathcal{O}(10^{12})}{particle/s} and survive several years in such an environment.
The next section describes how a neutrino beam line could be designed to keep the particle rate within the tracker capabilities.

\subsection{Beam Line}
\label{sec:1.2}

Three handles are available to reduce the beam particle rate. First, the particles can be spread in time by extracting them from the accelerator over a few seconds instead of the few micro-seconds cycle conventionally used. Second, the particles can be spread in space by adapting the beam transverse profile. Last, the particles can be momentum selected to keep only the \Pi's that would produce neutrinos in an energy range relevant for the phenomena under study.

While reducing the \Pi\ rate, the slow extraction is also preventing the use of magnetic pulsed horns traditionally employed to collimate the \Pi's. These elements could be replaced with quadrupoles~\cite{hep-ph_CareyEtAl_1971}. The ENUBET collaboration has recently demonstrated that quadrupoles sets~\cite{hep-ph_TortiEtAl_2020} can effectively reach a focusing power comparable to those of horns. Moreover, the quadrupoles can be arranged to focus both \PiP\ and \PiM. While this feature is considered to be problematic for conventional beams, it is clearly desired for a \tagged\ beam where the neutrino chirality is determined event-by-event.

% \begin{figure*}[h!]
% \includegraphics[width=\textwidth]{figs/beamline}
% \caption{Schematics of a possible beam line enabling neutrino \tagging. The schematics is described in the text.}
% \label{fig:beamline}
% \end{figure*}
% \noindent%

Based on these considerations, a beam line design, as shown in \autoref{fig:beamline}, could be envisaged. In this design, the protons are brought onto the target over few seconds using a slow extraction. The charged particles emerging from the target are refocused using four quadrupoles to ensure similar acceptances for \PiP's and \PiM's. Then, the particles are momentum-selected by a dipole magnet and a collimator. This momentum selection is expected to reduce the particle rate by one to two orders of magnitude by removing the low momentum charged particles~\cite{hep-ph_Pavlovic_2008}. The beam is split into two branches by the dipole. The positively charged particles are deflected in one direction and the negatively charged one in the opposite direction. In each branch, the beam particles are restored on trajectories parallel to the initial ones by a dipole magnet with magnetic field opposite to the first one. Finally, the same arrangement of magnets, but placed in reversed order, restore the beam particles on trajectories aligned with the initial ones. The four magnets are thus forming an achromat. Two sets of time-resolved tracking stations are installed inside and after the achromat. They allow to measure the direction of the \Pi\ as the particle trajectories inside and outside the achromat are parallel. The momentum is obtained by measuring the displacement between the two trajectories which scales with the particle rigidity.
The \Pi's then traverse a \SI{\mathcal{O}(100)}{m} long beam pipe where they may decay. At the end of the decay pipe, a dipole magnet with two sets of tracking stations, one after and one before the magnet, allow to measure the \Mu\ direction, electric charge and momentum.
% \renewcommand{\bottomfraction}{.5}

% \begin{strip}% to keep image and caption on one page
% 	\makebox[\linewidth]{%        to center the image
% 		\includegraphics[width=\textwidth]{figs/beamline}
% 	}
% 	\captionof{figure}{Schematic of a possible beam line enabling neutrino \tagging. Blue rectangles represent quadrupoles, red triangles dipoles and vertical dotted lines correspond to tracking planes. The number and location of those are not optimised. The schematic is described in more details in the text.}\label{fig:beamline}%      only if needed  
% \end{strip}
% \noindent

The beam line section upstream of the decay tube entrance has to be as short as possible as \PiMuNu\ decays occurring in this place cannot be reconstructed. Fortunately, the amount of neutrinos from early decays that happen to be in the far detector acceptance are significantly reduced by the improper collimation of the \Pi\ beam up to the last quadrupole.
Likewise, after the last dipole magnet and tracking plane, the particles should be stopped as quickly as possible to prevent untrackable \Pi\ and \Mu\ decays.

The \Pi\ rates shown in \autoref{fig:beamline} are derived assuming the capabilities of the HL-LHC trackers and a beam transverse size of at least \SI{0.1}{m^2}. The neutrino rate is derived from this value assuming that the \Pi\ momentum is \SI{\mathcal{O}(1-10)}{GeV/c} and the beam pipe is \SI{\mathcal{O}(100)}{m} long.

\subsection{\textit{Interacting} and \textit{Tagged} Neutrinos Association}
\label{sec:1.3}

The \tagging\ technique relies on the unambiguous matching between the \interacting\ neutrino and \tagged\ neutrino. This matching is performed based on time and angular coincidences.

The \tagged\ neutrino time coordinate will be determined with great precision, as each pixel layer will provide an independent time measurement with \SI{\mathcal{O}(10)}{ps} resolution. Hence the size of the matching time window will be determined, in the first place, by the resolution on the \interacting\ neutrino. The latter is typically \SI{\mathcal{O}(10)}{ns}.
Given this value and a neutrino flux of \SI{10^{11}}{\nu/s}, about \SI{\mathcal{O}(10^3)}{} \tagged\ neutrinos will coincide in time with a given \interacting\ neutrino.

The number of accidentally matching \tagged\ neutrinos will further be reduced by using the angular coincidence between them and the \interacting\ neutrino. The efficiency to reduce the number of matches is determined by the resolutions on the \tagged\ and \interacting\ neutrinos' directions.
%The coincidence is more effectively reducing the number of matching \tagged\ neutrinos for larger beam divergences. Thus, a conservative working hypothesis consists in assuming that the \Pi\ beam is perfectly focused and that the \NuANuMu\ beam divergence arises only from the \PiMuNu\ decay which is typical $1/\gamma$, where $\gamma$ is the \Pi\ boost.

The direction of the \interacting\ neutrino can be derived as the ratio of the transverse position of the neutrino interaction to the baseline. The resolution on the interaction position depends on the technology used for the neutrino detector and ranges from meters, for the sparsest instruments~\cite{hep-ph_KM3Net_2016}, to millimeters for the densest ones~\cite{hep-ph_DUNE_2020,hep-ph_Super-Kamiokande_2003}. Using the most spatially resolved detectors technology for SBNE's, and coarser ones for LBNE's, one can always achieve angular resolutions better than \SI{\mathcal{O}(10)}{\micro rad}.

\begin{figure}[t]
	\centering\includegraphics[width=0.8\columnwidth]{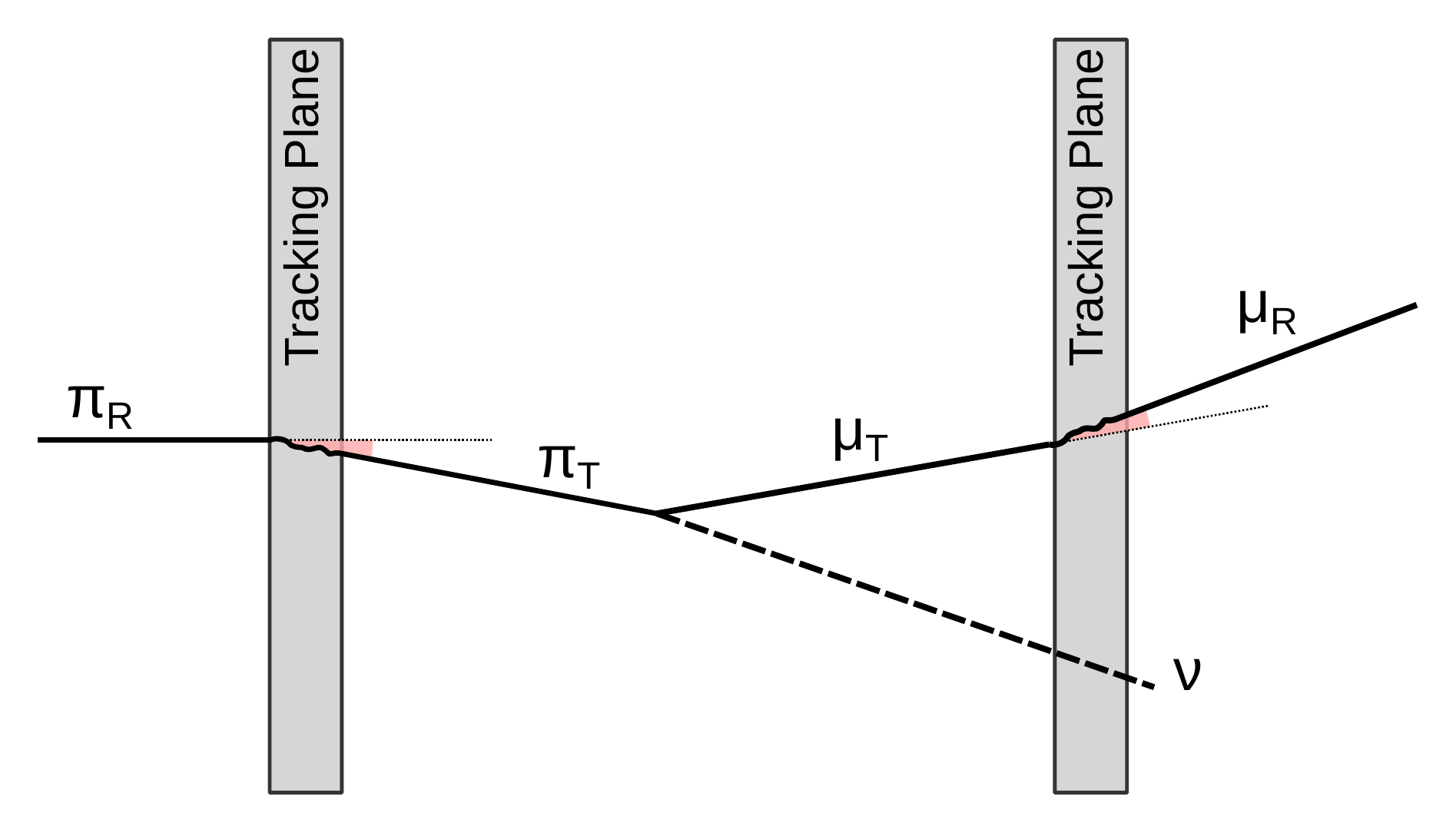}
	\caption{Schematic describing the hypothesis made on the achievable resolutions on \Pi\ and \Mu\ direction. At the decay point, the reconstructed \Pi\ and \Mu\ directions, $\pi_R$ and $\mu_R$, differ from the true ones, $\pi_T$ and $\mu_T$, as the \Pi\ and \Mu\ undergo multiple coulomb scattering in the last and first tracking plane they respectively cross.}
	\label{fig:hypoReso}
\end{figure}

The angular resolution on the \tagged\ neutrino is determined by the performances of the beam spectrometers. In this study, these performances are assumed to be similar to the ones of the existing NA62 GigaTracKer~\cite{hep-ph_AglieriRinellaEtAl_2019}. The momentum resolution is 0.2\% for the \Pi\ and \Mu. The resolution on the \Pi\ and \Mu\ direction is limited by the multiple coulomb scattering that the \Pi\ and \Mu\ undergo, respectively, in the last and first tracking plane they crossed, as illustrated in \autoref{fig:hypoReso}. The tracking planes are assumed to have a thickness of 0.5\% of a radiation length as for the NA62-GigaTraKer~\cite{hep-ph_AglieriRinellaEtAl_2019}.

\begin{figure}[b] 
	\includegraphics[width=\columnwidth]{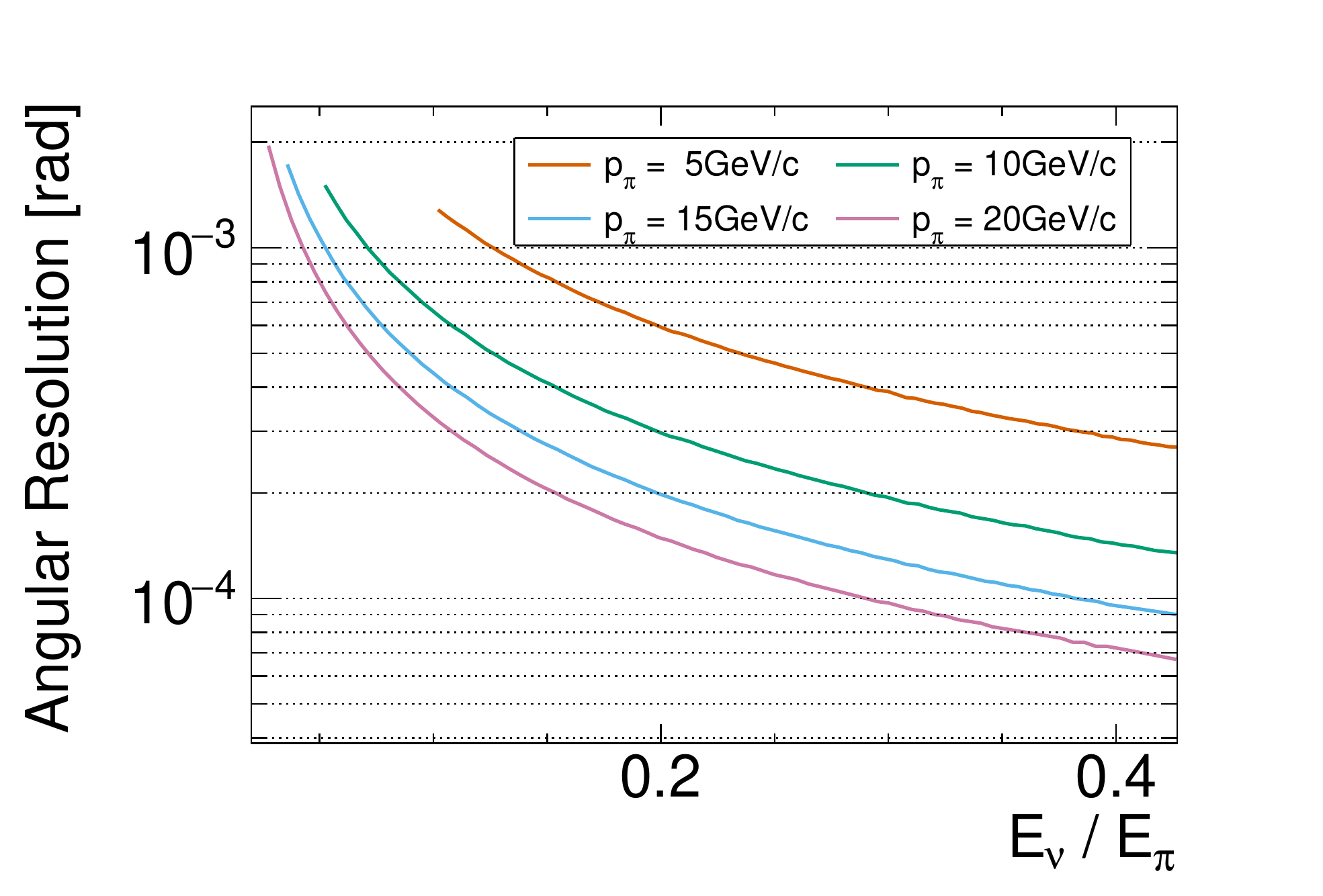}
	\caption{Angular resolution on the \tagged\ neutrino as a function of the fraction of the \Pi\ energy transferred to the neutrino for different \Pi\ momenta.
	%The top horizontal axis reports the tangent of the angle between the \NuMu\ and the \Pi\ multiplied by the \Pi\ Lorentz boost which is respectively 36, 72, 108 and 144 for a \Pi\ of 5, 10, 15 and \SI{25}{GeV/c}.
	}
	\label{fig:Reso}
\end{figure}

Under these hypotheses, the standard deviation on the space angle between the true and the reconstructed neutrino is shown in \autoref{fig:Reso} for different incoming \Pi\ momenta and as a function of the neutrino energy\footnote{As a reminder, the neutrino energy is uniformly distributed between $E_\nu/E_\pi = 0$ and $0.43$.}.
The best resolutions are achieved for high momentum \Pi's and high energy \NuANuMu's which are emitted colinear to the \Pi's. On average, the \tagged\ neutrino angular resolution ranges between 0.1 and \SI{1}{mrad}. These values are one to two orders of magnitude worse than the angular resolutions obtained for the \interacting\ neutrinos. Hence, the capability to correctly associate \tagged\ and \interacting\ neutrinos, based on the angular coincidence, is determined in the first place by the \tagged\ neutrino resolution.

% \noindent%
% \begin{minipage}{\linewidth}% to keep image and caption on one page
% 	\makebox[\linewidth]{%  
% 		\includegraphics[width=\columnwidth]{figs/AngularReso.pdf}
% 	}
% 	\captionof{figure}{Angular resolution on the \tagged\ neutrino (reconstructed kinematically from the \Pi\ and \MuP) as a function of the fraction of the \Pi\ energy transferred to the neutrino for different \Pi\ momenta.
% 		%The top horizontal axis reports the tangent of the angle between the \NuMu\ and the \Pi\ multiplied by the \Pi\ Lorentz boost which is respectively 36, 72, 108 and 144 for a \Pi\ of 5, 10, 15 and \SI{25}{GeV/c}.
% 	}
% 	\label{fig:Reso}
% \end{minipage}
% \noindent%
% \newpage

% \begin{strip}% to keep image and caption on one page
% 	\begin{minipage}{0.5\textwidth}
% 		\includegraphics[width=\textwidth]{figs/Distribution}
% 		\centering(a)
% 		\label{sfig:Distribution}
% 	\end{minipage}
% 	\begin{minipage}{0.5\textwidth}
% 		\includegraphics[width=\textwidth]{figs/CumulativeDistribution}
% 		\centering(b)
% 		\label{sfig:CumulativeDistribution}
% 	\end{minipage}
% 	\captionof{figure}{(a) Distribution of the angle between the neutrino and \Pi\ for different \Pi\ momenta and (b) the corresponding cumulative distributions. The width of the curves in (b) corresponds to the angular resolution on the \tagged\ neutrino direction. The black circled dots are the points at which the \tagged\ neutrino direction is no longer compatible with \SI{0}{rad}.}
% 	\label{fig:Distro}
% \end{strip}

This resolution has to be compared with the \tagged\ neutrino angular distribution. The wider this distribution, the smaller the number of accidentally matching \tagged\ neutrinos. Hence, a conservative hypothesis consists in assuming that the \Pi\ beam is perfectly focused and that the \NuANuMu\ beam divergence arises only from the \PiMuNu\ decay. In these conditions, the \NuANuMu\ beam divergence is around $1/\gamma$, where $\gamma$ is the \Pi\ Lorentz boost. For \Pi's with a momentum of \SI{15}{GeV/c}, the beam divergence is \SI{\Oo{10}}{mrad} while the angular resolution is about \SI{\Oo{0.1}}{mrad}. As a result, the number of accidentally matching \tagged\ neutrinos will be reduced by a factor $(0.1/10)^2$, going from 1000 to 0.1. As the worsening of the resolution at lower \Pi\ momenta is compensated by the increase of the beam divergence, the previous result is expected to be independent of the \Pi\ momentum. Assuming that the number of accidentally matching \tagged\ neutrinos follows a Poisson distribution\footnote{This hypothesis may be optimistic as time structures in the beam may remain due to improper de-bunching.}, the association between \interacting\ and \tagged\ neutrino will be unambiguous for 90\% of the events.

\begin{figure}[h!]
	\centering
	\includegraphics[width=0.99\columnwidth]{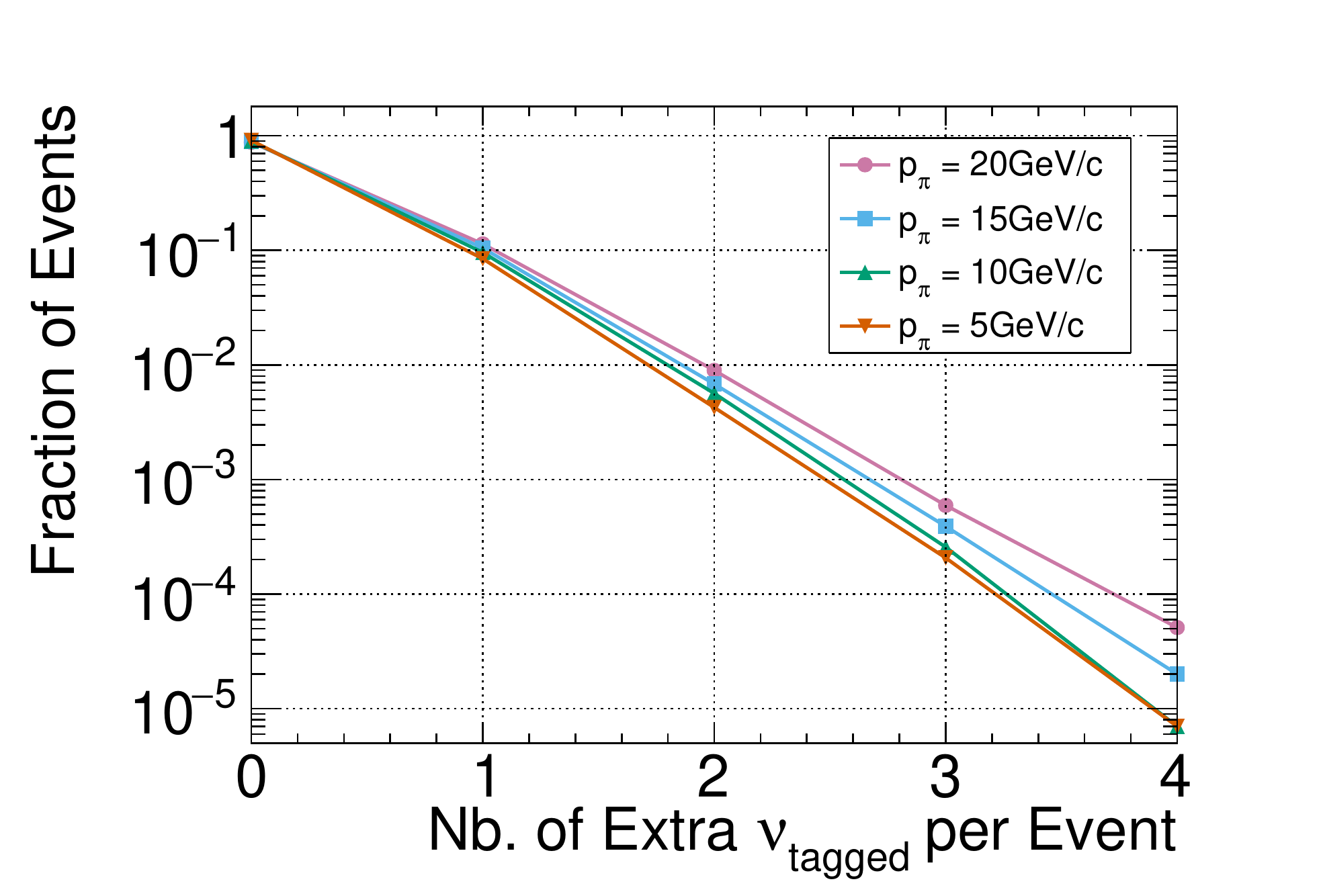}
	\caption{Distribution of the number of extra \tagged\ neutrinos in coincidence with the \interacting\ neutrinos assuming different \Pi\ momenta.}
	\label{fig:CoinRate}
\end{figure}

More quantitative simulations have confirmed these qualitative results, as reported in \autoref{fig:CoinRate}. According to the study, \tagged\ and \interacting\ neutrinos can be associated without ambiguity for more than 90\% of the events with a very marginal dependence on the \Pi\ momentum. The remaining 10\% of the events would have to be discarded for physics analyses as the association is ambiguous. Background from mis-associated events would only occur if the true \tagged\ neutrino is not reconstructed. Such a situation occurs when the \Pi's decay before the trackers. These early decays represent the main source of missing \tagged\ neutrinos. Compared to it, other sources, like tracking inefficiencies, can be made negligible. Assuming that the fraction of \interacting\ neutrinos originating from early decays is of the order of \Oo{1}\%, the probability for a mis-tagged event is thus \Oo{0.1}\%.
Hence, these results, obtained with conservative hypotheses, indicate that neutrino \tagging\ in a beam with a rate of $10^{11}$ \NuANuMu/s should be feasible with the technologies developed for the HL-LHC.
% \noindent%
% \begin{minipage}{\linewidth}
% 	\makebox[\linewidth]{%  
% 		\includegraphics[width=\columnwidth]{figs/Coincidence.pdf}
% 	}%
% 	\captionof{figure}{Distribution of the number of extra \tagged\ neutrinos in coincidence with the \interacting\ neutrinos for different \Pi\ momenta.}%
% 	\label{fig:CoinRate}%
% \end{minipage}
% \noindent%
%\newpage

\subsection{Energy Resolution}
\label{sec:1.5}

As mentioned in \autoref{sec:benefits}, the neutrino energy resolution obtained from the kinematical reconstruction is expected to greatly surpass the ones obtained from neutrino detectors in the GeV energy range. The neutrino energy, $E_\nu$, can be derived from the \Pi\ momentum, $p_\pi$ and the angle between the \Pi\ and \NuANuMu, $\theta_{\pi\nu}$  as
\begin{equation}
	E_\nu = \frac{(1-m^2_\mu/m^2_\pi)p_\pi}{1+\gamma^2\theta_{\pi\nu}^2},
\end{equation}
where $\gamma$ is the \Pi\ Lorentz boost.

\begin{figure}[h!]
	\centering\includegraphics[width=.99\columnwidth]{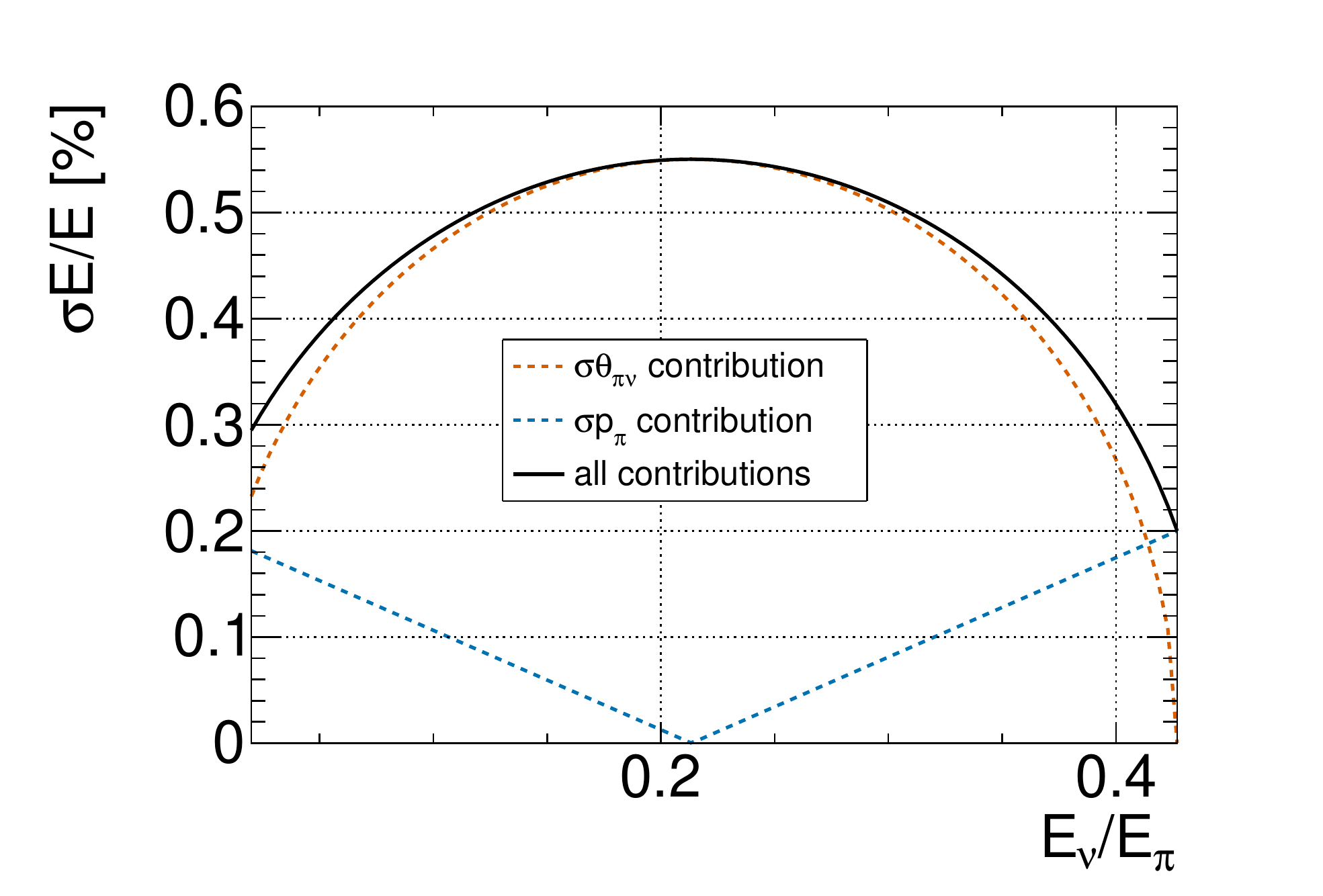}
	\caption{Energy resolution on the \tagged\ neutrino as a function of the fraction of the \Pi\ energy transferred to the $\nu$ (black solid) overlaid with the contributions from the resolutions on the angle between the \Pi\ and \Nu\ (red dashed) and on the \Pi\ momentum (blue dashed). 
	%The resolutions are independent of the \Pi\ momentum.
	}
	\label{fig:EReso}
\end{figure}

The uncertainties on neutrino direction can be assumed to be negligible (see \autoref{sec:1.3}). Hence the uncertainty on $\theta_{\pi\nu}$ is dominated by the multiple coulomb scattering of the \Pi\ in the last tracking plane it crossed. Assuming a momentum resolution on the \Pi\ similar to what is achieved at NA62~\cite{hep-ph_AglieriRinellaEtAl_2019,hep-ph_NA62_2017c} ($\sigma_p/p = 0.2\%$), the neutrino energy resolution is expected to range between 0.6\% and 0.2\% as shown in \autoref{fig:EReso}, and is independent of the \Pi\ momentum.

% \begin{equation}
% 2\cdot \theta_{MCS} \cdot  \frac{E_\nu}{E_\pi} \frac{m_\pi}{m_\pi^2-m_\mu^2} \sqrt{\frac{E_\pi}{E_\nu}\frac{m_\pi^2-m_\mu^2}{m_\pi^2}   - 1} 
% \end{equation}

% 100. * 13.6/1000*sqrt(0.005)*(1+0.038*TMath::Log(0.005)) * fabs( x * 0.13957018/(0.13957018*0.13957018-0.105658*0.105658) * 2 * TMath::Sqrt( (0.13957018*0.13957018-0.105658*0.105658) / x / 0.13957018/0.13957018 - 1) )

\subsection{Prospects for an Experimental Demonstrator}
\label{sec:1.6}
In the short term, the NA62 experiment should be able to demonstrate the feasibility of the neutrino \tagging\ technique. The collaboration is aiming at collecting about $10^{13}$ \SI{75}{GeV/c} kaon decays~\cite{hep-ph_NA62Collaboration_2010,hep-ph_NA62_2017c} and most of them are \KMuNu. Given the size of the NA62 liquid krypton calorimeter (\SI{20}{ton})~\cite{hep-ph_NA48_2002}, few hundreds of \NuMu\ should interact in the krypton. These events could then be matched with the \Kp\ and \MuP\ reconstructed in the NA62 spectrometers. Since 2021, a trigger line dedicated to these events has been operational.

% \section{Tagged Short Base Line Neutrino Experiment}
% \label{sec:3}
% \input{TSBL}

\section{\textit{Tagged} Long Baseline Neutrino Experiments}
\label{sec:3}

\subsection{A New Paradigm}

The next generation of LBNE's will be devoted to the precision measurement of the neutrino oscillation parameters and in particular the CP violating phase $\delta_{CP}$. These measurements require both large neutrino samples and small systematic uncertainties.  In this context, the \tagging\ technique would be very advantagous as it would greatly reduce the systematic uncertainties as explained in \autoref{sec:benefits}.

However, the limitation imposed by the \tagging\ on the beam particle rate prevents to use this method for the new generation of experiments, DUNE~\cite{hep-ph_DUNE_2020,hep-ph_DUNE_2020a,hep-ph_DUNE_2020b} and T2HK~\cite{hep-ph_Hyper-Kamiokande_2018,hep-ph_Hyper-Kamiokande_2021}, as they are relying on beams of very high intensity to collect enough statistics.

A \tagged\ LBNE would thus need a very large detector to collect enough neutrinos with a modest beam intensity. An interesting option is to use natural water Cherenkov neutrino detectors such as KM3NeT/ORCA~\cite{hep-ph_KM3Net_2016}. Compared to most of the neutrino telescopes like ANTARES~\cite{hep-ph_ANTARES_2011}, KM3NeT/ARCA~\cite{hep-ph_KM3Net_2016}, IceCube~\cite{hep-ph_IceCube_2017} or Baikal-GVD~\cite{hep-ph_AvrorinEtAl_2019}, which are primarily dedicated to neutrinos with energy above \SI{1}{TeV}, KM3NeT/ORCA specifically aims at studying the oscillations of atmospheric neutrinos in the energy range between 3 and \SI{100}{GeV}. Using this technology, very large volumes of water can be instrumented for reasonable costs, as no excavation is required. For example KM3NeT/ORCA will instrument around \SI{6.8}{Mton} of sea water, i.e. a number of scattering centres more than a hundred time larger than the ones of DUNE~\cite{hep-ph_DUNE_2020,hep-ph_DUNE_2020a,hep-ph_DUNE_2020b} or HK~\cite{hep-ph_Hyper-Kamiokande_2018,hep-ph_Hyper-Kamiokande_2021}.

While being less granular and precise than these two detectors, this technology should be sufficient for a \tagged\ LBNE. Indeed, the initial properties of each neutrino being measured with an unprecedented precision, the detector is mainly left with the identification of the flavour of the oscillated neutrinos.

Hence, a LBNE with a \tagged\ beam and a mega-ton scale natural water neutrino detector should provide in about ten years of operation a sample of \Oo{10^5} neutrinos~\cite{hep-ph_AkindinovEtAl_2019} of the highest quality with very small systematic uncertainties. This option is therefore a viable solution for the next generations of LBNE's. In the next sections a case study of such an experiment from the U70 accelerator complex in Protvino, Russia, to KM3NeT/ORCA is presented. Note that similar LBNE's could be implemented between U70 and lake Baikal in Russia or between Fermilab and the Neptune submarine infrastructure offshore of British Columbia~\cite{hep-ph_Vallee_2016}.

\subsection{A \tagged\ LBNE from Protvino to KM3NeT/ORCA}

\begin{figure*}[h!]
	\centering
	\begin{subfigure}[b]{0.49\textwidth}
		\centering
		\includegraphics[width=\textwidth]{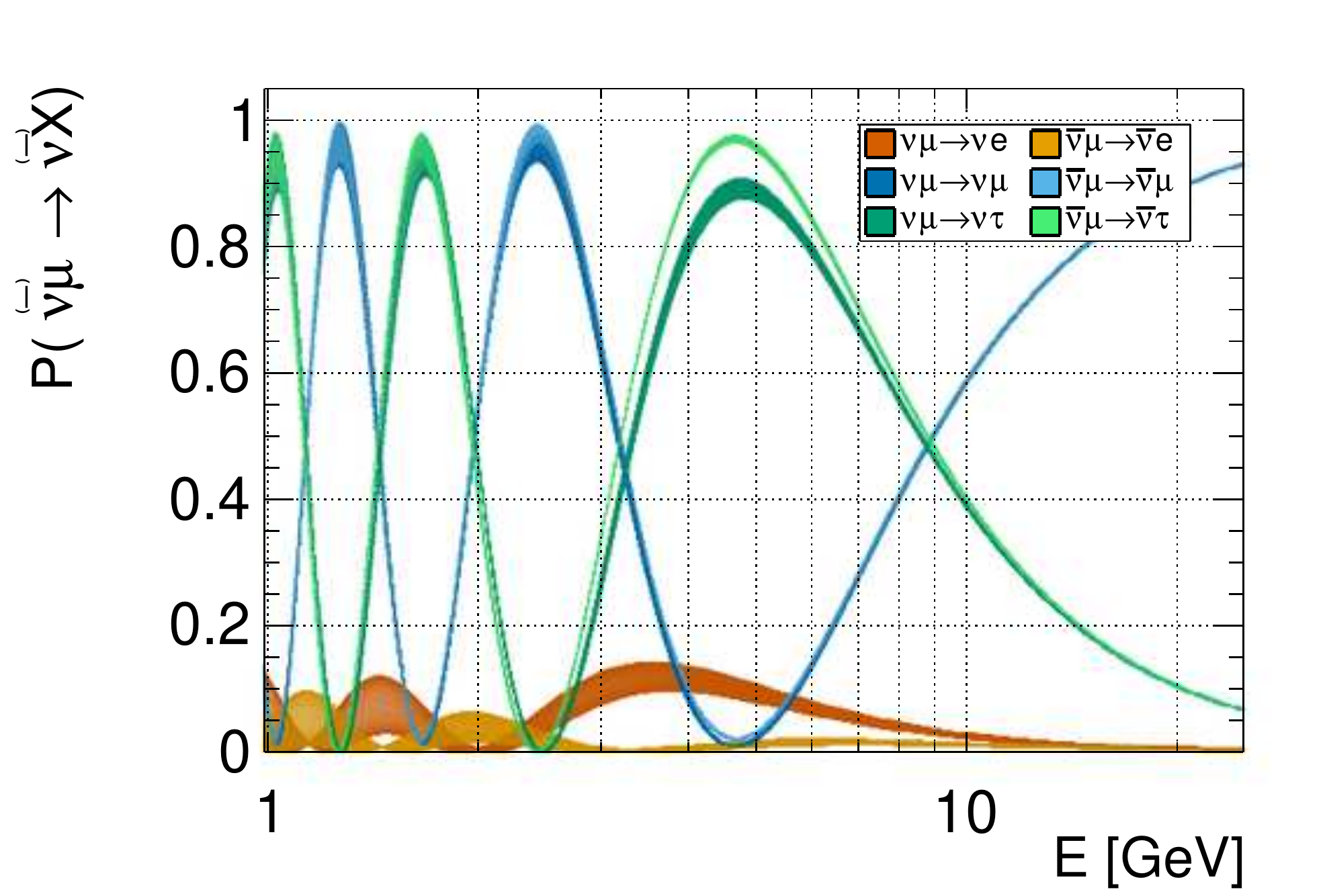}
		\caption{}
		\label{fig:OscAll}
	\end{subfigure}
	\hfill
	\begin{subfigure}[b]{0.49\textwidth}
		\centering
		\includegraphics[width=\textwidth]{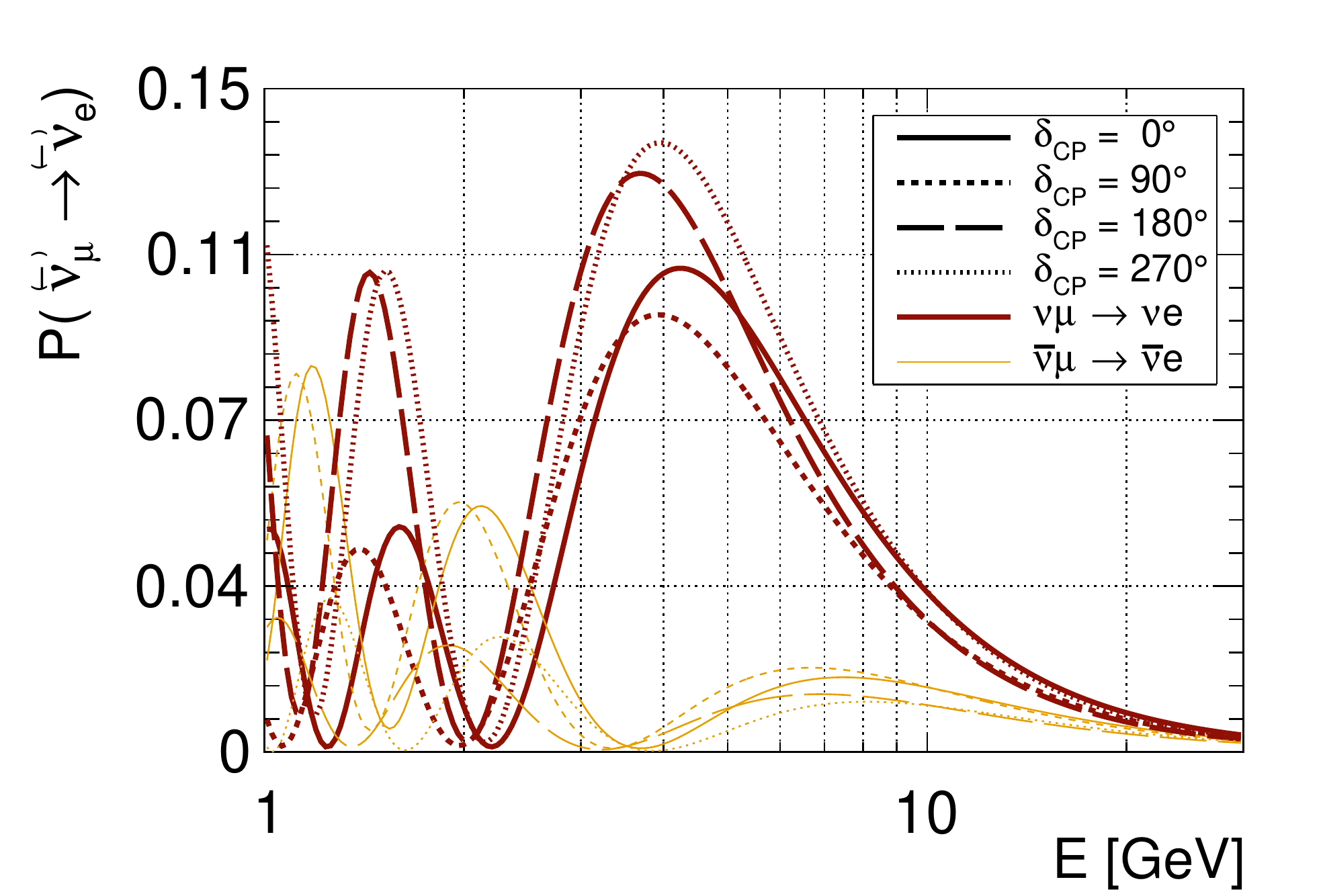}
		\caption{}
		\label{fig:P2OOscProb}
	\end{subfigure}
	\caption{ (\protect\subref{fig:OscAll}) Probabilities for \NuMu's (dark colour scatter plots) and \ANuMu's (light colour scatter plots) to oscillate to each neutrino flavours as function of the neutrino energy and for all possible values of $\delta_{CP}$. (\protect\subref{fig:P2OOscProb}) Probabilities for a \NuMu's (dark red thick lines) and \ANuMu's (yellow thin lines) to oscillate to the electron neutrino flavour as a function of the neutrino energy. The probabilities for different $\delta_{CP}$ values are shown with different line styles. In both (\protect\subref{fig:OscAll}) and  (\protect\subref{fig:P2OOscProb}), the oscillation baseline is \SI{2595}{km}. The oscillation probabilities are computed with the OscProb software package~\cite{OscProb} and using the oscillation parameters from~\cite{hep-ph_EstebanEtAl_2020}.}
	\label{fig:proba}
\end{figure*}

The KM3NeT/ORCA detector is under construction offshore Toulon, France and the first detection lines deployed already allowed to observe the oscillation of atmospheric neutrinos~\cite{hep-ph_KM3NeT_2021a}.
The possibility to perform a LBNE from the U70 accelerator complex in Protvino, Russia, to KM3NeT/ORCA was discussed in detail in~\cite{hep-ph_AkindinovEtAl_2019}. The experiment is referred to as P2O. The baseline of \SI{2595}{km} corresponds to an energy at the first oscillation maximum of around ~\SI{5}{GeV}, as shown in~\autoref{fig:proba}, which is well above the detection threshold of KM3NeT/ORCA. In the following paragraphs, a study of the sensitivity to $\delta_{CP}$ of P2O with a \tagged\ beam is presented.

The study assumes that a \SI{450}{kW} wide band beam can be delivered by U70~\cite{hep-ph_Omega}. The neutrino rates and spectra are assumed to be identical to the ones obtained in the initial P2O study~\cite{hep-ph_AkindinovEtAl_2019}\footnote{Studies should be performed to refine the hypotheses made on the neutrino rates and spectra as the performances of a tagged beam line can significantly differs from the ones in [32]}. Such a beam would allow to collect about $20\cdot10^3$ neutrinos and $5\cdot10^3$ anti-neutrinos per year with KM3NeT/ORCA. The beam power corresponds to \SI{2.25\cdot 10^{14}}{protons~per~pulse}~\cite{hep-ph_Omega}. The same order of magnitude is expected for the \Pi\ rate after the protons interacted in the target and before any selection. This rate can be reduced by around two orders of magnitude by imposing a minimum \Pi\ momentum of \SI{9}{GeV/c}~\cite{hep-ph_Pavlovic_2008}. As the maximum neutrino energy from \PiMuNu\ is $0.43\cdot E_\pi$, this selection has no effect on the neutrino with an energy around and above \SI{5}{GeV}, the first oscillation maximum. The expected \Pi\ rate should thus be around \SI{10^{12}}{particle/s} which is within the capabilities of the trackers as discussed in \autoref{sec:1.1}.

For what concerns the association between \interacting\ and \tagged\ neutrinos, the resolution on the \interacting\ neutrino time-of-flight will be dominated by the uncertainties on the interaction position. The later is expected to be \SI{1}{m}~\cite{hep-ph_KM3Net_2016} corresponding to about \SI{3}{ns} which is better than the value assumed in \autoref{sec:1.3}. The individual association of the \interacting\ neutrino with the \tagged\ one can thus be taken as granted.

\subsection{Measurement Principle}
\begin{figure*}
	\centering
	\begin{subfigure}[b]{0.49\textwidth}
		\centering
		\includegraphics[width=\textwidth]{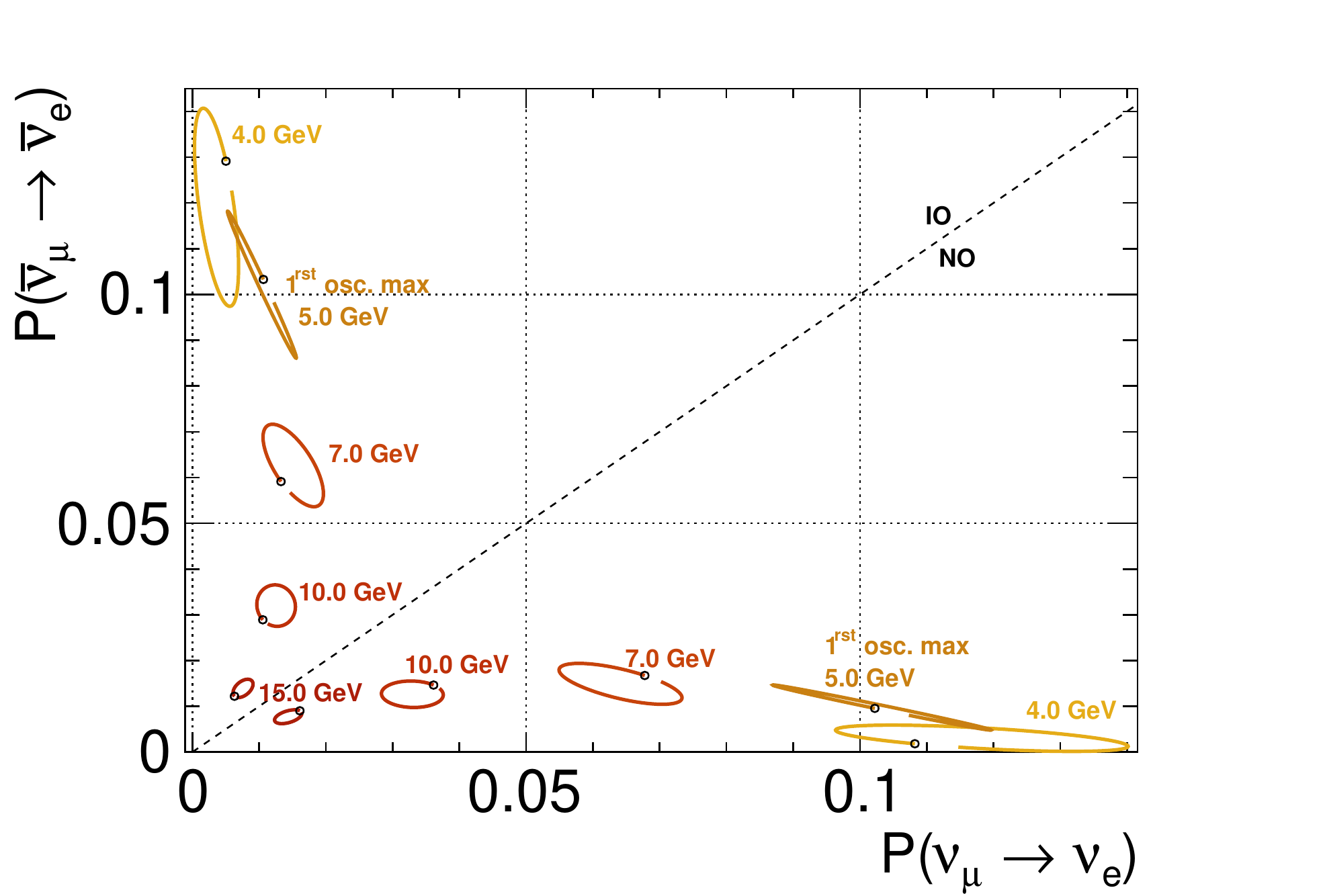}
		\caption{}
		\label{fig:BiProba}
	\end{subfigure}
	\hfill
	\begin{subfigure}[b]{0.49\textwidth}
		\centering
		\includegraphics[width=\textwidth]{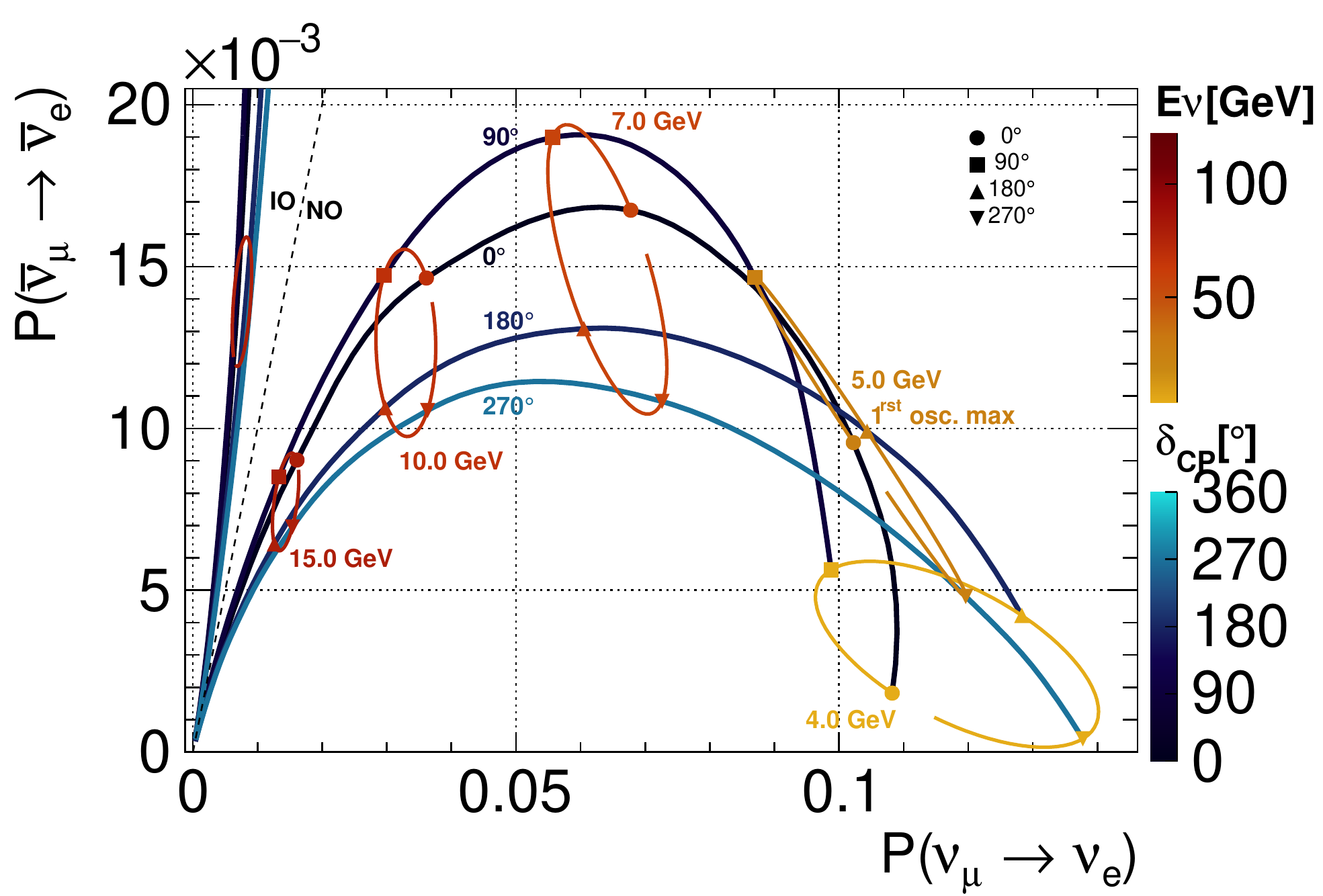}
		\caption{}
		\label{fig:ProbEDcp}
	\end{subfigure}
	\caption{
		(\protect\subref{fig:BiProba}) Probability for \ANuMu\ to oscillate to \ANuE\ versus the probability for \NuMu\ to oscillate to \NuE. For each neutrino energy, the two probabilities describe an ellipse as $\delta_{CP}$ is varied. The black circles indicates the points for which $\delta_{CP}$ equals \SI{0}{\degree}. The color of the ellipses corresponds to the neutrino energy and is reported on the red color scale in (\protect\subref{fig:ProbEDcp}). The ellipses obtained assuming normal ordering (NO) and inverted ordering (IO) are clearly separated with IO above the $P(\NuMu\to\NuE)=P(\ANuMu\to\ANuE)$ line (dashed line) and NO below. A zoom in the NO region is shown in  (\protect\subref{fig:ProbEDcp}). The blue lines represents the probabilities for a given $\delta_{CP}$ value when the energy is varied. At high energy both probabilities are null as no oscillation occurs. In both (\protect\subref{fig:BiProba}) and  (\protect\subref{fig:ProbEDcp}), the oscillation baseline is \SI{2595}{km}. The probabilities are computed with the OscProb software package~\cite{OscProb} and using the oscillation parameters from~\cite{hep-ph_EstebanEtAl_2020}.}
	\label{fig:BiProbs}
\end{figure*}

The \tagged\ P2O experimental setup will access an unprecedented neutrino energy resolution which opens new possibilities. The standard method to determine $\delta_{CP}$~\cite{hep-ph_T2K_2020} consists in measuring the probabilities for neutrino and anti-neutrino oscillation, $\rm{P}(\NuMu \to \NuE)$ and $\rm{P}(\ANuMu \to \ANuE)$, and in comparing them to the expectations. The latters describe two ellipses, one for each mass ordering, in the $\rm{P}(\NuMu \to \NuE) \times \rm{P}(\ANuMu \to \ANuE)$ plane. At \tagged\ P2O, the excellent energy resolution allows to extend the method and to measure the two probabilities for different energies.
% Hence instead of confronting the mesured point to the two expected ellipses,  one for each mass ordering a sets of points of ellipses can be tested.
\autoref{fig:BiProbs} shows these two probabilities as function of $\delta_{CP}$ and for various energies between 4 and \SI{15}{GeV}. The probabilities are nearly symmetric with respect to the $\rm{P}(\NuMu \to \NuE) = \rm{P}(\ANuMu \to \ANuE)$ line. The top part corresponds to inverted ordering (IO) and the bottom to normal ordering (NO). With such a long baseline, the two orderings are well separated. \autoref{fig:BiProbs}(\protect\subref{fig:ProbEDcp}) shows a zoom into the NO region. For each energy, the points corresponding to the different $\delta_{CP}$ values describe an ellipse. The points corresponding to the same $\delta_{CP}$ value follow, as the energy is varied, one of the curved lines in shades of blue.
At high energy (dark red ellipses), both probabilities are null, as the oscillation is no longer occurring. Near the first oscillation maximum energy, \SI{5}{GeV}, the curvature of the ellipse is maximal at $\delta_{CP} = 90^\circ$ and $180^\circ$ which translates in the well known result that the precision to measure $\delta_{CP}$ is the worse at these values. However, the ellipses apsides correspond to other $\delta_{CP}$ values for other energies. In addition, at these energies the ellipses are more circular. The excellent energy reconstruction offered by the \tagging\ technique makes it possible to resolve the different ellipses. Hence, the degradation of the $\delta_{CP}$ precision at $90^\circ$ and $180^\circ$ is expected to be much less pronounced at a \tagged\ P2O.

% \begin{figure*}[!ht]
%     \includegraphics[width=\columnwidth]{figs/BiProba.pdf}
%  	\includegraphics[width=\columnwidth]{figs/ProbEDcp.pdf}
% 	\caption{Expected values for the oscillation probabilties $\rm{P}(\NuMu \to \NuE)$ and $\rm{P}(\ANuMu \to \ANuE)$. The probabilities are nearly symmetric with respect to the identity (dashed line) and so the plot focuses on the bottom part. The top part corresponds to the inverse neutrino mass ordering and the bottom to the normal ordering. Each ellipse corresponds to probabilities for the different $\delta_{CP}$ values and for a fixed neutrino energy. The energy is reported as the line color (top color scale). The points for $\delta_{CP} = 0, 90, 180~\rm{and}~270^\circ$ are reported by four different markers: a circle, a square, an up pointing and a down pointing triangles.  The points on the ellipses corresponding to the same $\delta_{CP}$ values form, as the energy is varied, curly lines. These lines are shown in shades of blues and the corresponding $\delta_{CP}$ value is reported on the bottom color scale.}
% 	\label{fig:ProbEDcp}
% \end{figure*}

\subsection{Detector Responses}
The performances of the KM3NeT/ORCA detector in terms of energy response, effective mass and particle identification (PID) are assumed to be identical to the ones obtained on atmospheric neutrinos. A detailed description of the performances is available in~\cite{hep-ph_KM3NeT_2021}.
This hypothesis is conservative as new reconstruction and triggering algorithms could be developed to exploit the fact that the direction and energy of the beam neutrinos are known \textit{a-priori}. A second scenario is also considered in the study where the detector photo-cathode density is assumed to be twice as large as the KM3NeT/ORCA nominal value. In this case, the performances for a given energy are assumed to be equal to those obtained at KM3NeT/ORCA for twice the energy. A third limit case scenario is also envisaged where the PID is assumed to be perfect. The energy resolution on the \tagged\ neutrino is assumed to be 1\% which is also a conservative hypothesis.

\subsection{\textit{Tagged} P2O Sensitivity to $\delta_{CP}$}
With the assumptions described above, the sensitivity of \tagged\ P2O to $\delta_{CP}$ is derived with a method similar to the one described in~\cite[Sect.~3.1]{hep-ph_KM3NeT_2021} and using the oscillation parameters from~\cite{hep-ph_EstebanEtAl_2020}.
%% a bit more details?
The analysis is performed using the OscProb~\cite{OscProb} and ROOT~\cite{hep-ph_BrunEtAl_1997} software packages. The neutrino and anti-neutrino data samples are analysed in the plane made by the energy reconstructed by the tagger and the one reconstructed by the KM3NeT/ORCA detector. Three event categories are considered based on the detector PID response: a track-like class collecting mostly \NuANuMu-CC and \NuANuTau-CC where the $\tau^\pm$ decay to a \Mu; a shower-like class collecting mostly \NuANuE-CC, NC and \NuANuTau-CC where the $\tau^\pm$ decay hadronically; and an intermediate class collecting an admixture of flavours. A full description of the KM3NeT/ORCA PID performances is available in~\cite[Fig.~6]{hep-ph_KM3NeT_2021}.
When a perfect PID is considered, four event categories are used, one for each flavour and one for the NC interaction. In this case, the analysis is performed in one dimension corresponding to the \tagged\ neutrino reconstructed energy.
The distributions of the energy reconstructed by KM3NeT/ORCA and by the tagger in each PID category is obtained by applying the detector response to the true energy distributions.

Several systematic uncertainties, reported in \autoref{tab:syst} are included in the model to reflect the limited knowledge on:
\begin{itemize}
	\item the oscillation parameters,  $\theta_{13}$, $\theta_{23}$ and $\dmsq{23}$,
	\item the detector performances in terms of detection efficiency, energy scale and PID,
	\item the beam neutrino rates,
	\item the cross-section.
\end{itemize}
The choice and treatment of these uncertainties are similar to what is described in~\cite{hep-ph_AkindinovEtAl_2019} and in~\cite[Sect.~3.1]{hep-ph_KM3NeT_2021}. Technically, the systematic uncertainties are implemented with a set of energy scales and re-normalisation factors which distort the expected event distributions.

A first energy scale, referred to as \textit{global energy scale} in \autoref{tab:syst}, is applied to neutrinos from all channels (\NuANuE, \NuANuMu\ and \NuANuTau\ both CC and NC) and represents the uncertainties on the detector energy response which originate from the limited knowledge on the photo-detection efficiency. The large fluctuations of the hadronic showers light yield further increase the uncertainties on the detection efficiency. Hence, a second scaling factor, referred to as \textit{hadronic energy scale} in \autoref{tab:syst}, is applied to neutrinos from all channels but weighted by the average fraction of light produced by the hadronic shower in the interaction. Finally, as the energy thresholds and the energy responses may differ between the channels, a third factor, referred to as \textit{\NuANuE$_{,\mu}$ energy scale} in \autoref{tab:syst}, is applied only to \NuANuE$_{,\mu}$-CC, as in~\cite{hep-ph_AkindinovEtAl_2019}.

The uncertainties on the PID response are implemented with a set of independent energy scale factors, referred to as \textit{PID category energy scales} in \autoref{tab:syst}, and applied to the neutrinos in each PID category. In addition, the event yields classified in each PID category are scaled by independent re-normalisation factors, referred to as \textit{PID category normalisations} in \autoref{tab:syst}. These re-normalisation factors reflect also the uncertainties on the total event yield which originate from the limited knowledge on the detection efficiency, beam rate and cross-sections.

As the cross sections for NC and \NuANuTau-CC interactions are less precisely known than the others, two independent re-normalisation factors, \textit{NC cross section} and \textit{\NuANuTau-CC cross section} in \autoref{tab:syst}, are respectively applied to the event yields from NC and \NuANuTau-CC.
% Several systematic uncertainties are included to reflect the limited knowledge on:
% 
% \begin{itemize}
% 	\item the oscillation parameters,  $\theta_{13}$, $\theta_{23}$ and $\dmsq{23}$,
% 	\item the detector response in terms of energy scale and particle identification,
% 	\item the normalisation of the event rates,
% 	\item the \NuANuTau-CC, \NuANuE$_{,\mu}$-CC and NC absolute cross sections.
% \end{itemize}
% 
% Technically, three energy scales are considered. The first one applies to all channels (\NuANuE, \NuANuMu and \NuANuTau both CC and NC) and represents the uncertainties on the photo-detection efficiency. A second scaling is weighted by the fraction of the light arising from the hadronic shower. A third one applies only to \NuANuE$_{,\mu}$-CC events. The uncertainties on the PID response is implemented as an energy scale and a re-normalisation of the event rate classified in each PID category. The latter reflects also the uncertainties on the total number of events.

Gaussian priors are applied to these parameters. The standard deviations of these priors are identical to \cite{hep-ph_AkindinovEtAl_2019} and reported in \autoref{tab:syst}.

% \noindent
% \begin{minipage}{\textwidth} 
\begin{table}[!h]
	\centering
	\begin{tabular}{rl}\hline
		Parameter                     & Gaussian Prior Std Dev   \\\hline
		$\theta_{13}$                 & $0.15^\circ$             \\
		$\theta_{23}$                 & $2.0^\circ$              \\
		$\dmsq{23}$                   & \SI{5\cdot10^{-3}}{eV^2} \\
		Global energy scale           & 3\%                      \\
		Hadronic energy scale         & 3\%                      \\
		\NuANuE$_{,\mu}$ energy scale & 3\%                      \\
		PID category energy scales    & 3\%                      \\
		PID category normalisations    & 10\%                     \\
		\NuANuTau-CC cross section    & 10\%                     \\
		\NuANuE$_{,\mu}$ cross section  & 10\%                     \\
		NC cross section              & 5\%                      \\
		\hline
	\end{tabular}
	\caption{Parameters considered as systematic uncertainties together with the standard deviation of the Gaussian priors applied to them.}
	\label{tab:syst}
\end{table}
% \end{minipage}

\begin{figure}[t]
	\includegraphics[width=\columnwidth]{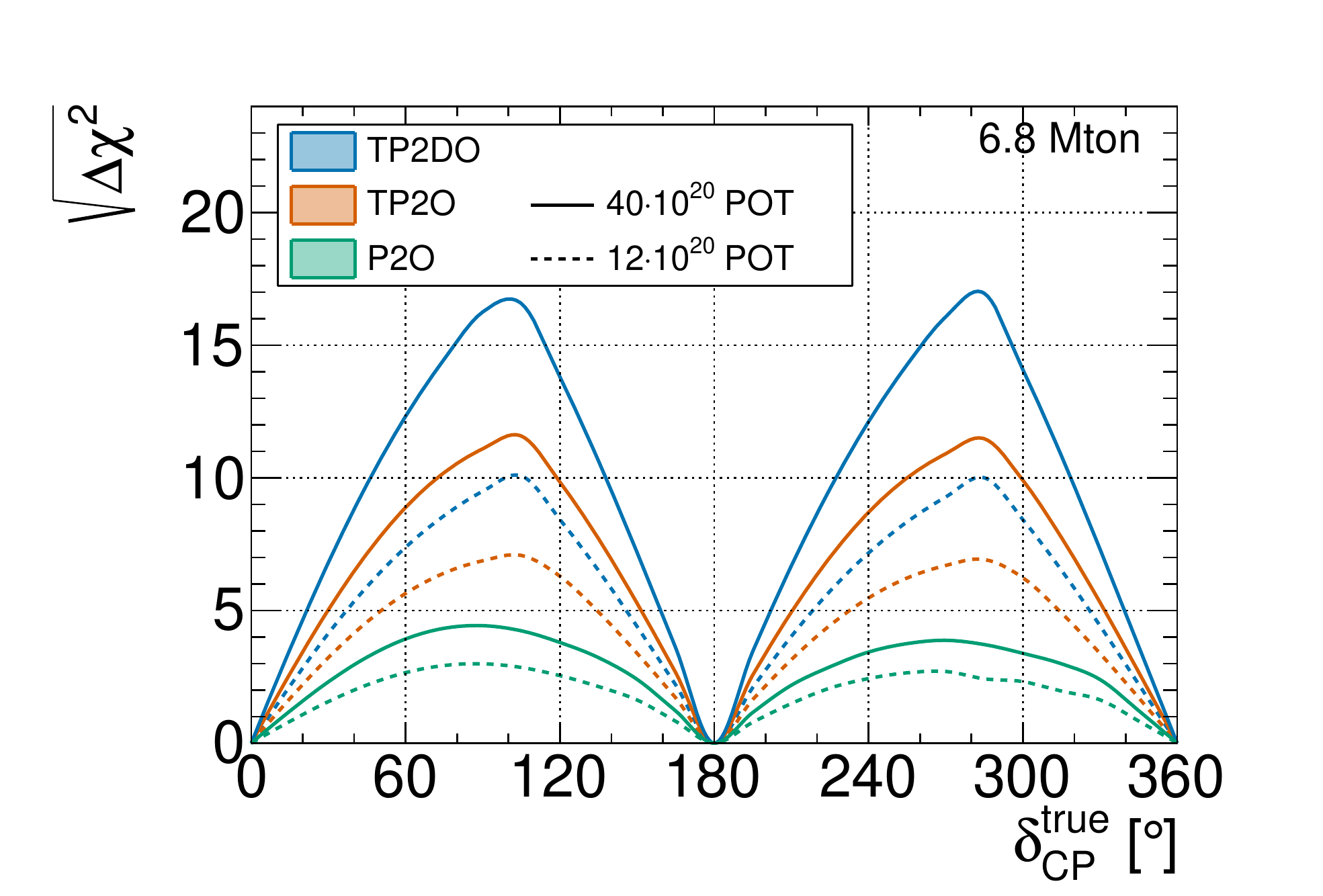}
	\caption{Sensitivity to exclude the no-CP violation hypothesis as a function of the true $\delta_{CP}$ value for P2O (green), \tagged\ P2O (red) and \tagged\ P2O with a far detector with a photocathode density twice as large as KM3NeT/ORCA (blue). The solid lines correspond to an exposure of \SI{40\cdot10^{20}}{POT} and the dashed ones to \SI{12\cdot10^{20}}{POT}.}
	\label{fig:cpv}
\end{figure}

The sensitivity to exclude the CP-conservation hypothesis is reported in \autoref{fig:cpv} for different scenarios: standard P2O, \tagged\ P2O and \tagged\ P2O with a denser detector, and, for two different exposures: \si{12\cdot10^{20}} protons-on-target (POT) and \SI{40\cdot10^{20}}{POT} corresponding to 3 and 10 years of operation with a \SI{450}{kW} beam. Discovering the CP violation in the neutrino sector appears to be impossible at P2O. However such a discovery becomes possible with the \tagging\ technique. With \SI{12\cdot10^{20}}{POT}, \tagged\ P2O would be able to claim a \SI{5}{\sigma} discovery of this effect for 46\% of the $\delta_{CP}$ phases violating the CP symmetry and 68\% with \SI{40\cdot10^{20}}{POT}. These values are increased to 60\% and 76\% if a denser detector is used.

\begin{figure}[t]
	\includegraphics[width=\columnwidth]{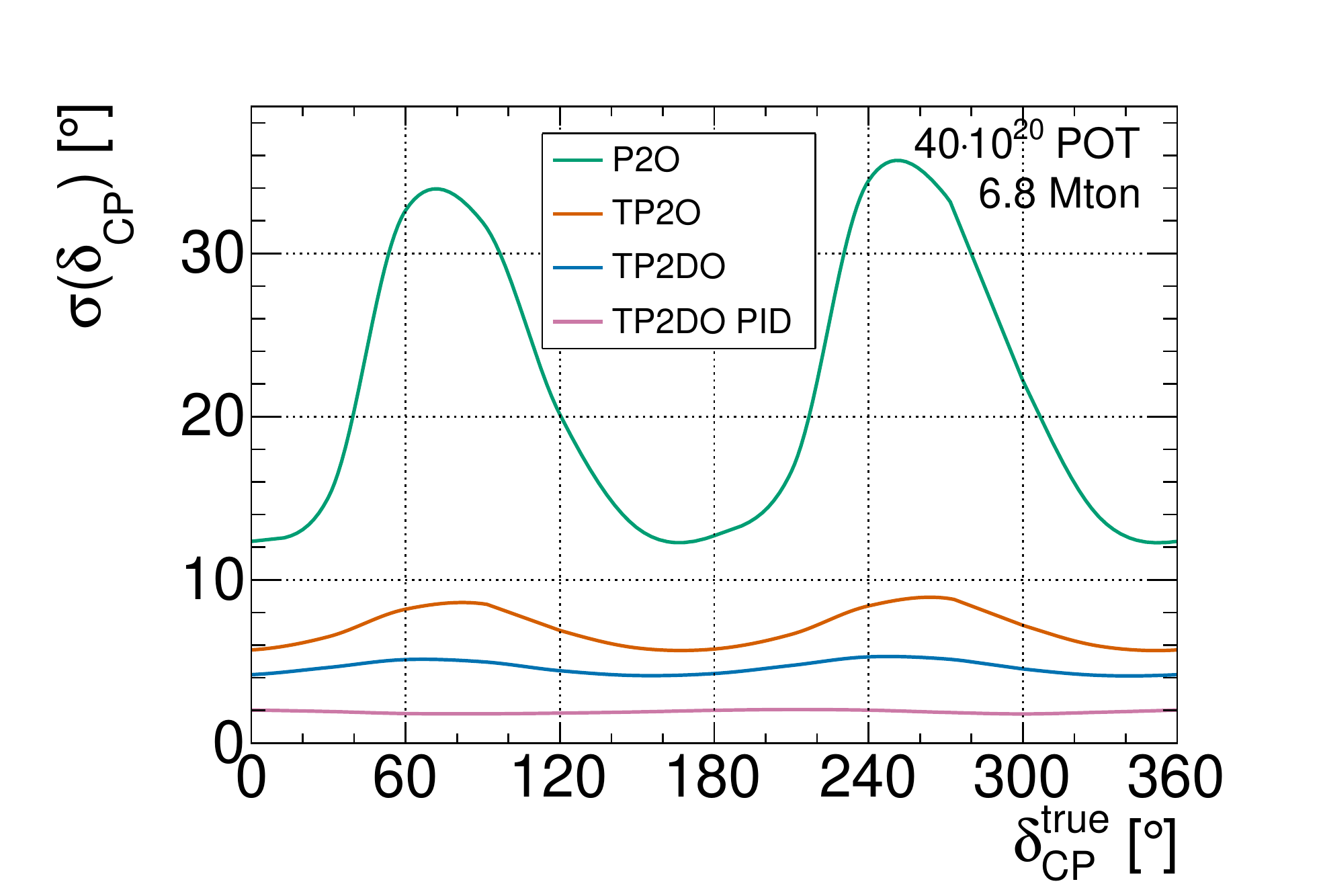}
	\caption{Precision on $\delta_{CP}$ as a function of the true $\delta_{CP}$ value for P2O (green), \tagged\ P2O (red), \tagged\ P2O with a far detector with a photocathode density twice as large as KM3NeT/ORCA (blue), and with a perfect PID (purple).}
	\label{fig:dcp}
\end{figure}

The precision on $\delta_{CP}$ is reported in \autoref{fig:dcp} for four scenarios: standard P2O, \tagged\ P2O, \tagged\ P2O with a denser detector and, finally, with a perfect PID. The benefit of the \tagging\ method is very clear. It allows to reach a much better precision and the precision obtained remains stable over the whole $\delta_{CP}$ range. In the case of \tagged\ P2O with a dense detector, a precision between \SI{4}{\degree} to \SI{5}{\degree} is expected for an exposure of \SI{40\cdot10^{20}}{POT} and a water instrumented mass of \SI{6.8}{Mton}. In the limit case for which a perfect PID is achieved, a \SI{2}{\degree} precision could be reached.
% \newpage
% Finally, the $\delta_{CP}$ precision as a function of the exposure is represented in \autoref{fig:dcptime}. The precision is given for the most ($\delta_{CP}$=\SI{0}{\degree}) and worse ($\delta_{CP}$=\SI{270}{\degree}) favourable true $\delta_{CP}$ values. The exposure is normalised to the number of POT and to the mass of instrumented water as the KM3NeT technology opens the possibility to instrument much larger volume for reasonable costs. As a key figure, the precision expected for
% KM3NeT/ORCA with \SI{4\cdot10^{20}}{POT}, corresponding to one year with a \SI{450}{kW} beam, is expected to be between \SI{25}{\degree} and \SI{14}{\degree} which is the precision foreseen at the next generation of LBNE~\cite{hep-ph_DUNE_2020,hep-ph_Itow_2019} after 7 years of operation.

% \begin{figure}[!ht]
% 	\includegraphics[width=\columnwidth]{figs/PrecTime.pdf}
% 	\caption{Precision on $\delta_{CP}$ as a function of exposure assuming a true $\delta_{CP}$ of \SI{0}{\degree} (solid lines) and \SI{270}{\degree} (dashed) for \tagged\ P2O (red) and \tagged\ P2O with a far detector with a photocathode density twice as large as ORCA (blue).}
% 	\label{fig:dcptime}
% \end{figure}

%Short baseline:
% -superb energy resolution
% -flux below 1\%
% -angular correlation

\
\section{Conclusions and Prospects}
\label{sec:conclusion}
In this article, a new experimental method was presented for accelerator based neutrino experiments: the neutrino \tagging. The method consists in exploiting the neutrino production mechanism, the \PiMuNu\ decay, to kinematically reconstruct the neutrino properties based on the incoming and outgoing decay charged particles. The  reconstruction of these particles relies on the recent progress and on-going developments in silicon particle detector technology which can operate at very high particle flux.
The \tagging\ method allows to reconstruct individually nearly all neutrinos in the beam and to determine the particle properties with an unprecedented precision. Using time and angular coincidences, the neutrino \interacting\ in the detector can be individually matched to the \PiMuNu\ decay it originated from and to the corresponding \tagged\ neutrino.

The benefits brought by this method are numerous. Such a precise knowledge of the neutrino source allows to drastically reduce the systematic uncertainties and background contaminations for neutrino oscillation studies. These studies also benefit from the excellent energy resolution which allows to fully exploit the energy dependence of the oscillation probabilities. Finally, the \tagging\ technique enables significant improvements of the cross-section measurements and of the
phenomenological models used to infer the neutrino energy from the neutrino–nucleus interactions.
%%%% FIXE-ME
% Moreover, these studies will further indirectly benefit from the \tagging\ technique as it would enable significant improvements of the cross-section measurements and of the
% phenomenological models which are used to infer the neutrino energy from the neutrino–nucleus interactions.

The implementation of this technique requires to design neutrino beam lines where the particle flux remains within the capabilities of the silicon detector technologies. Ideas were presented on how to design such a beam line using slow extraction, large beam transverse size and careful momentum selection of secondary pions. The resulting beam line layout employs only basic and affordable elements such as dipoles and quadrupoles. The beam line can simultaneously collect neutrinos and anti-neutrinos by exploiting the event-by-event chirality determination provided by the \tagging. A generic beam line design is under investigation within the CERN Physics Beyond Colliders Study Group~\cite{pbc} and in collaboration with the Institute for High Energy Physics in Protvino. The outcome of these studies will allow to refine the hypotheses made in this article.

Based on these ideas, a new type of long baseline neutrino experiments was proposed which uses a \tagged\ beam together with a mega-ton scale natural water Cherenkov neutrino detector.
The strength of the design stems from the counterbalancing of the detector limitations by the assets of the \tagging\ and vice-versa. Indeed, the coarseness of the sparse water Cherenkov detector is overcome by the excellent precision of the \tagging\ and, conversely, the rate limitation imposed by the \tagging\ is outweighed by the size of the detector.

% The implementation of such an experiment could be enviseaged in several places for example between Fermilab and the NEPTUNE Ocean Observatory~\cite{}, or between Protvino and lake Baikal or KM3NeT-ORCA. The latter was used
The physics potential for this new type of long baseline experiments was evaluated with the Protvino to KM3NeT/ORCA (P2O) setup as a benchmark. The reduced systematic uncertainties and the sub-percent energy resolution yields unprecedented sensitivities to the CP violating phase $\delta_{CP}$. 
% If the CP symmetry is violated, \tagged\ P2O would be able in 10 years to claim a discovery of this phenomena for 68\% of the CP-violating phases. Moreover the \tagging\ opens the possibility to measure the $\delta_{CP}$ phase with a precision of few degrees.
Several scenarios were considered for the far detector performances extrapolating from the ones obtained with atmospheric neutrinos. These scenarios still require to be consolidated with precise studies of the KM3NeT/ORCA detector performances with \tagged\ beam neutrinos. 
% Dedicated studies should be carried out to refine these hypotheses.
The most optimistic scenarios indicate that a \SI{2}{\degree} precision on $\delta_{CP}$ could be achieved. Hence, the \tagging\ technique represents a valuable option for the next generations of neutrino experiments. More investigations will be carried out to study the complete physics case of the method at short and long baseline experiments.

\begin{acknowledgements}
The study presented in this article was done under the auspices of the Centre de Physique des Particules de Marseille and the Agence Nationale de la Recherche through the ANR-19-CE31-0009 grant. The author is grateful to V. Dabhi for the discussions related to tagged neutrino angular resolution, to N. Charitonidis and E. Parozzi for their insights on beam line design, to J. Brunner for his enlightening views in numerous topics related to the neutrino oscillation, to C. Vall\'ee for his support and valuable feedback on the \textit{tagging} idea, and to D. Dornic for reviewing the manuscript of this article. The author is also grateful to the KM3NeT Collaboration for providing information related to the KM3NeT/ORCA detector performances, to the CERN's PBC Working Group for offering a forum to discuss new ideas and for supporting their development, to the NRC Kurchatov Institutes for discussing and studying the possibility to implement a \textit{tagged} beam between Protvino and KM3NeT/ORCA and, to the NA62 Collaboration for offering the possibility to experimentally demonstrate the feasibility of the \textit{tagging} method.

\end{acknowledgements}
% \newpage
\bibliographystyle{spphys}       % APS-like style for physics
\bibliography{ref}   % name your BibTeX data base

\begin{thebibliography}{10}
\providecommand{\url}[1]{{#1}}
\providecommand{\urlprefix}{URL }
\expandafter\ifx\csname urlstyle\endcsname\relax
  \providecommand{\doi}[1]{DOI \discretionary{}{}{}#1}\else
  \providecommand{\doi}{DOI \discretionary{}{}{}\begingroup
  \urlstyle{rm}\Url}\fi

\bibitem{hep-ph_DUNE_2020}
B.~Abi, et~al., Eur. Phys. J. C \textbf{80}(10), 978 (2020).
\newblock \doi{10.1140/epjc/s10052-020-08456-z}

\bibitem{hep-ph_DUNE_2020a}
B.~Abi, et~al., Deep underground neutrino experiment (dune), far detector
  technical design report, volume ii: Dune physics.
\newblock Tech. rep. (2020)

\bibitem{hep-ph_DUNE_2020b}
B.~Abi, et~al., JINST \textbf{15}(08), T08008 (2020).
\newblock \doi{10.1088/1748-0221/15/08/T08008}

\bibitem{hep-ph_Hyper-Kamiokande_2018}
K.~Abe, et~al.
\newblock {Hyper-Kamiokande Design Report}.
\newblock \url{https://arxiv.org/abs/1805.04163} (2018)

\bibitem{hep-ph_Hyper-Kamiokande_2021}
Y.~Itow, PoS \textbf{ICRC2021}, 1192 (2021).
\newblock \doi{10.22323/1.395.1192}

\bibitem{hep-ph_Pontecorvo_1979}
B.~Pontecorvo, Lett. Nuovo Cim. \textbf{25}, 257 (1979).
\newblock \doi{10.1007/BF02813638}

\bibitem{hep-ph_Nedyalkov_1984}
I.P. Nedyalkov.
\newblock {Single spectrometer station for neutrino tagging}.
\newblock \url{http://inis.jinr.ru/sl/NTBLIB/JINR-E1-84-515.pdf} (1984)

\bibitem{hep-ph_Bohm_1987}
G.~Bohm.
\newblock Project of a tagged neutrino facility at serpukhov.
\newblock
  \url{https://inis.iaea.org/collection/NCLCollectionStore/_Public/21/060/21060898.pdf}
  (1987)

\bibitem{hep-ph_BernsteinEtAl_1990}
R.H. Bernstein, et~al.
\newblock {A Proposal for a Neutrino Oscillation Experiment in a Tagged
  Neutrino Line}.
\newblock \url{https://inspirehep.net/files/94bcaf83ca2fdca7476a57e6182e4868}
  (1990)

\bibitem{hep-ph_AnikeevEtAl_1998}
V.B. Anikeev, et~al., Nucl. Instrum. Meth. A \textbf{419}, 596 (1998).
\newblock \doi{10.1016/S0168-9002(98)00837-7}

\bibitem{hep-ph_AglieriRinellaEtAl_2019}
G.~Aglieri~Rinella, et~al., JINST \textbf{14}, P07010 (2019).
\newblock \doi{10.1088/1748-0221/14/07/P07010}

\bibitem{hep-ph_Lai_2018}
A.~Lai,  (IEEE, Sydney, NSW, Australia, 2018), pp. 1--3.
\newblock \doi{10.1109/NSSMIC.2018.8824310}

\bibitem{hep-ph_SadrozinskiEtAl_2013}
H.F.W. Sadrozinski, et~al., Nucl. Instrum. Meth. A \textbf{730}, 226 (2013).
\newblock \doi{10.1016/j.nima.2013.06.033}

\bibitem{hep-ph_HuberEtAl_2008}
P.~Huber, et~al., JHEP \textbf{03}, 021 (2008).
\newblock \doi{10.1088/1126-6708/2008/03/021}

\bibitem{hep-ph_BrancaEtAl_2021}
A.~Branca, G.~Brunetti, A.~Longhin, M.~Martini, F.~Pupilli, F.~Terranova,
  Symmetry \textbf{13}(9), 1625 (2021).
\newblock \doi{10.3390/sym13091625}

\bibitem{hep-ph_LSND_2001}
A.~Aguilar-Arevalo, et~al., Phys. Rev. D \textbf{64}, 112007 (2001).
\newblock \doi{10.1103/PhysRevD.64.112007}

\bibitem{hep-ph_MiniBooNE_2013}
A.~Aguilar-Arevalo, et~al., Phys. Rev. Lett. \textbf{110}, 161801 (2013).
\newblock \doi{10.1103/PhysRevLett.110.161801}

\bibitem{hep-ph_LudoviciEtAl_1996}
L.~Ludovici, P.~Zucchelli.
\newblock {Conceptual study of an antitagged experiment searching for $\nu_\mu
  \to \mu_e$ oscillation}.
\newblock \url{https://arxiv.org/abs/hep-ex/9701007}

\bibitem{hep-ph_BoothEtAl_2019}
A.C. Booth, et~al., Phys. Rev. Accel. Beams \textbf{22}(6), 061003 (2019).
\newblock \doi{10.1103/PhysRevAccelBeams.22.061003}

\bibitem{hep-ph_Adrian-MartinezEtAl_2017}
S.~Adri\'an-Mart\'\i{}nez, et~al., JHEP \textbf{05}, 008 (2017).
\newblock \doi{10.1007/JHEP05(2017)008}

\bibitem{hep-ph_FriedlandEtAl_2019}
A.~Friedland, S.W. Li, Phys. Rev. D \textbf{99}(3), 036009 (2019).
\newblock \doi{10.1103/PhysRevD.99.036009}

\bibitem{hep-ph_AnkowskiEtAl_2015}
A.M. Ankowski, et~al., Phys. Rev. D \textbf{92}(9), 091301 (2015).
\newblock \doi{10.1103/PhysRevD.92.091301}

\bibitem{hep-ph_CLASe4v_2021}
M.~Khachatryan, et~al., Nature \textbf{599}(7886), 565 (2021).
\newblock \doi{10.1038/s41586-021-04046-5}

\bibitem{hep-ph_ParkeEtAl_2016}
S.~Parke, et~al., Phys. Rev. D \textbf{93}(11), 113009 (2016).
\newblock \doi{10.1103/PhysRevD.93.113009}

\bibitem{hep-ph_NA62Collaboration_2010}
F.~Hahn, et~al., {NA62: Technical Design Document}.
\newblock Tech. rep., CERN, Geneva (2010).
\newblock \urlprefix\url{https://cds.cern.ch/record/1404985}

\bibitem{hep-ph_NA62_2017c}
E.~Cortina~Gil, et~al., JINST \textbf{12}(05), P05025 (2017).
\newblock \doi{10.1088/1748-0221/12/05/P05025}

\bibitem{hep-ph_LHCbCollaboration_2017}
R.~Aaij, et~al., {Expression of Interest for a Phase-II LHCb Upgrade:
  Opportunities in flavour physics, and beyond, in the HL-LHC era}.
\newblock Tech. rep. (2017).
\newblock \urlprefix\url{https://cds.cern.ch/record/2244311}

\bibitem{hep-ph_LHCbVELOGroup_2021}
M.~van Beuzekom, JPS Conf. Proc. \textbf{34}, 010014 (2021).
\newblock \doi{10.7566/JPSCP.34.010014}

\bibitem{hep-ph_CareyEtAl_1971}
D.C. Carey, et~al., IEEE Trans. Nucl. Sci. \textbf{18}, 755 (1971).
\newblock \doi{10.1109/TNS.1971.4326174}

\bibitem{hep-ph_TortiEtAl_2020}
M.~Torti, et~al., Int. J. Mod. Phys. A \textbf{35}(34n35), 2044017 (2020).
\newblock \doi{10.1142/S0217751X20440170}

\bibitem{hep-ph_Pavlovic_2008}
Z.~Pavlovic, Studies of the neutrino flux for the numi beam at fnal.
\newblock Ph.D. thesis, PhD Thesis, University of Texas at Austin (2008).
\newblock
  \urlprefix\url{https://lss.fnal.gov/archive/thesis/2000/fermilab-thesis-2008-59.pdf}

\bibitem{hep-ph_KM3Net_2016}
S.~Adrian-Martinez, et~al., J. Phys. \textbf{G43}(8), 084001 (2016).
\newblock \doi{10.1088/0954-3899/43/8/084001}

\bibitem{hep-ph_Super-Kamiokande_2003}
Y.~Fukuda, et~al., Nucl. Instrum. Meth. A \textbf{501}, 418 (2003).
\newblock \doi{10.1016/S0168-9002(03)00425-X}

\bibitem{hep-ph_NA48_2002}
M.~Jeitler, Nucl. Instrum. Meth. A \textbf{494}, 373 (2002).
\newblock \doi{10.1016/S0168-9002(02)01505-X}

\bibitem{hep-ph_ANTARES_2011}
M.~Ageron, et~al., Nucl. Instrum. Meth. A \textbf{656}, 11 (2011).
\newblock \doi{10.1016/j.nima.2011.06.103}

\bibitem{hep-ph_IceCube_2017}
M.G. Aartsen, et~al., JINST \textbf{12}(03), P03012 (2017).
\newblock \doi{10.1088/1748-0221/12/03/P03012}

\bibitem{hep-ph_AvrorinEtAl_2019}
A.D. Avrorin, et~al., Bull. Russ. Acad. Sci. Phys. \textbf{83}(8), 921 (2019).
\newblock \doi{10.3103/S1062873819080057}

\bibitem{hep-ph_AkindinovEtAl_2019}
A.V. Akindinov, et~al., Eur. Phys. J. C \textbf{79}(9), 758 (2019).
\newblock \doi{10.1140/epjc/s10052-019-7259-5}

\bibitem{hep-ph_Vallee_2016}
C.~Vallee.
\newblock Pacific neutrinos: Towards a high precision measurement of cp
  violation ?
\newblock \url{https://arxiv.org/abs/1610.08655}

\bibitem{hep-ph_KM3NeT_2021a}
L.~Nauta, et~al., PoS \textbf{ICRC2021}, 1123 (2021).
\newblock \doi{10.22323/1.395.1123}

\bibitem{OscProb}
J.~Coelho.
\newblock {OscProb}.
\newblock \url{https://github.com/joaoabcoelho/OscProb/}

\bibitem{hep-ph_EstebanEtAl_2020}
I.~Esteban, et~al., JHEP \textbf{09}, 178 (2020).
\newblock \doi{10.1007/JHEP09(2020)178}

\bibitem{hep-ph_Omega}
The omega project.
\newblock \url{http://www.ihep.su/files/OMEGA%20LOI.pdf}

\bibitem{hep-ph_T2K_2020}
K.~Abe, et~al., Nature \textbf{580}(7803), 339 (2020).
\newblock \doi{10.1038/s41586-020-2177-0}.
\newblock [Erratum: Nature 583, E16 (2020)]

\bibitem{hep-ph_KM3NeT_2021}
S.~Aiello, et~al., Eur. Phys. J. C \textbf{82}(1), 26 (2022).
\newblock \doi{10.1140/epjc/s10052-021-09893-0}.
\newblock \urlprefix\url{https://doi.org/10.1140/epjc/s10052-021-09893-0}

\bibitem{hep-ph_BrunEtAl_1997}
R.~Brun, F.~Rademakers, Nucl. Instrum. Meth. A \textbf{389}, 81 (1997).
\newblock \doi{10.1016/S0168-9002(97)00048-X}

\bibitem{pbc}
The physics beyond colliders study group.
\newblock \url{https://pbc.web.cern.ch/}

\end{thebibliography}

%\begin{thebibliography}{}
% etc
%\end{thebibliography}

\end{document}